\newif\iffigs\figstrue

\documentstyle[12pt]{article}

\iffigs
  \input epsf
\else
\fi

\batchmode
  \newfont{\footscrfont}{rsfs10}
  \newfont{\footbbbfont}{msbm10}
\errorstopmode

\newif\ifscrf\scrftrue
\ifx\footscrfont\nullfont
  \scrffalse
\fi

\newif\ifamsf\amsftrue
\ifx\footbbbfont\nullfont
  \amsffalse
\fi

\def\ppnumber{CLNS-94/1307}
\def\ppdate{November 1994}
\def\pplogo{\vbox{\kern-\headheight\kern -1pt
\halign{##&##\hfil\cr&{
\ppnumber}\cr\rule{0pt}{2.5ex}&\ppdate\cr}
}}

\makeatletter
\date{}
\def\dedicatory#1{\def\@date{\normalsize\it#1}}
\def\subjclass#1{\def\@thefnmark{}\@footnotetext{1991
    {\it Mathematics Subject Classification.} #1}}
\def\keywords#1{\def\@thefnmark{}\@footnotetext{
    {\it Key words and phrases.} #1}}

\def\ps@firstpage{\ps@empty \def\@oddhead{\hss\pplogo}%
  \let\@evenhead\@oddhead 
}
\def\maketitle{\par
 \begingroup
 \def\thefootnote{\fnsymbol{footnote}}
 \def\@makefnmark{\hbox
 to 0pt{$^{\@thefnmark}$\hss}}
 \if@twocolumn
 \twocolumn[\@maketitle]
 \else \newpage
 \global\@topnum\z@ \@maketitle \fi\thispagestyle{firstpage}\@thanks
 \endgroup
 \setcounter{footnote}{0}
 \let\maketitle\relax
 \let\@maketitle\relax
 \gdef\@thanks{}\gdef\@author{}\gdef\@title{}\let\thanks\relax}

\def\abstract{\if@twocolumn
\section*{Abstract}
\else \small
\begin{center}
{\bf ABSTRACT}
\end{center}
\quotation
\fi}

\newif\iffn\fnfalse

\@ifundefined{reset@font}{\let\reset@font\empty}{} 
\long\def\@footnotetext#1{\insert\footins{\reset@font\footnotesize
    \interlinepenalty\interfootnotelinepenalty
    \splittopskip\footnotesep
    \splitmaxdepth \dp\strutbox \floatingpenalty \@MM
    \hsize\columnwidth \@parboxrestore
   \edef\@currentlabel{\csname p@footnote\endcsname\@thefnmark}\@makefntext
    {\rule{\z@}{\footnotesep}\ignorespaces
      \fntrue#1\fnfalse\strut}}}
\makeatother

\ifamsf
  \newfont{\bigbbbfont}{msbm10 scaled\magstep2}
  \newfont{\bbbfont}{msbm10 scaled\magstep1}  
  \newfont{\smallbbbfont}{msbm8}
  \newfont{\tinybbbfont}{msbm6}
  \newfont{\smallfootbbbfont}{msbm7}
  \newfont{\tinyfootbbbfont}{msbm5}
\fi

\ifscrf
  \newfont{\scrfont}{rsfs10 scaled\magstep1}  
  \newfont{\smallscrfont}{rsfs7}
  \newfont{\tinyscrfont}{rsfs7}
  \newfont{\smallfootscrfont}{rsfs7}
  \newfont{\tinyfootscrfont}{rsfs7}
\fi

\ifamsf
  \newcommand{\Bbb}[1]{\iffn
      \mathchoice{\mbox{\footbbbfont #1}}{\mbox{\footbbbfont #1}}
      {\mbox{\smallfootbbbfont #1}}{\mbox{\tinyfootbbbfont #1}}\else
      \mathchoice{\mbox{\bbbfont #1}}{\mbox{\bbbfont #1}}
      {\mbox{\smallbbbfont #1}}{\mbox{\tinybbbfont #1}}\fi}
\else
  \def\bigbbbfont{\bf}
  \def\Bbb{\bf}
\fi

\ifscrf
  \newcommand{\Scr}[1]{\iffn
    \mathchoice{\mbox{\footscrfont #1}}{\mbox{\footscrfont #1}}
    {\mbox{\smallfootscrfont #1}}{\mbox{\tinyfootscrfont #1}}\else
    \mathchoice{\mbox{\scrfont #1}}{\mbox{\scrfont #1}}
    {\mbox{\smallscrfont #1}}{\mbox{\tinyscrfont #1}}\fi}
\else
  \def\Scr{\cal}
\fi

\def\operatorname#1{\mathop{\rm #1}\nolimits}
\def\C{{\Bbb C}}
\def\F{{\cal F}}
\def\O{{\cal O}}
\def\P{{\Bbb P}}

\def\R{{\Bbb R}}
\def\Z{{\Bbb Z}}

\def\Img{\operatorname{Im}}
\def\Rea{\operatorname{Re}}

\def\opeq#1{\advance\lineskip#1 \advance\baselineskip#1
	\advance\lineskiplimit#1}

\def\eqalign#1{\null\,\vcenter{\opeq{2.5\jot}\mathsurround=0pt
	\everycr={}\tabskip=0pt
	\halign{\strut\hfil$\displaystyle{##}$&$\displaystyle{{}##}$\hfil
	\crcr#1\crcr}}\,\null}

\def\sm{$\sigma$-model}

\def\CY{Calabi-Yau}

\def\cM{{\Scr M}}

\def\cD{{\Scr D}}

\def\cMc{{\hfuzz=100cm\hbox to 0pt{$\;\overline{\phantom{X}}$}\cM}}
\def\barcD{{\hfuzz=100cm\hbox to 0pt{$\;\overline{\phantom{X}}$}\cD}}

\def\ff#1#2{{\textstyle\frac{#1}{#2}}}
\def\F#1#2{{}_{#1}F_{#2}}

\def\RoR{$R\leftrightarrow\alpha^\prime/R$}
\def\normalord#1{\mathord{:}#1\mathord{:}}
\def\Gep#1{#1_{\hbox{\scriptsize Gep}}}
\def\th#1{$#1^{\hbox{\scriptsize\it th}}$}

\begin{document}
\setcounter{page}0
\title{\LARGE The Moduli Space of \\ $N=2$ Superconformal Field
Theories\\[10mm]
\insert\footins{\hbox to\hsize{\footnotesize
Lectures given at the Trieste summer school 1994.\hfil}}}
\author{
Paul S. Aspinwall\\[0.7cm]
\normalsize F.R.~Newman Lab.~of Nuclear Studies,\\
\normalsize Cornell University,\\
\normalsize Ithaca, NY 14853\\[10mm]
}

{\hfuzz=10cm\maketitle}

\def\Large{\large}
\def\LARGE{\large\bf}

\vskip 1.5cm
\vskip 1cm

\begin{abstract}

We review the structure of the moduli space of particular $N=(2,2)$
superconformal field theories. We restrict attention to those of
particular use in superstring compactification, namely those with
central charge $c=3d$ for some integer $d$ and whose NS fields have
integer $U(1)$ charge. The cases $d=1$, 2 and 3 are analyzed.
It is shown that in the case $d\geq3$ it is
important to use techniques of algebraic geometry rather than rely on
metric-based ideas. The phase structure of these moduli spaces is
discussed in some detail.

\end{abstract}

\vfil\break

\section{Introduction}		\label{s:intro}

As well as applications to statistical physics, conformal field
theory has proved to be a very powerful tool in string theory. In
particular, the ground state of a critical string corresponds to a
conformal field theory with a specific central charge. It is
of particular interest to classify all such ground states which can
therefore be done by
finding the space of all conformal field theories of a given central
charge. This ``moduli space'' forms the space of string vacua and may
be considered as the stringy analogue of the space of Einstein metrics
in general relativity.

The moduli space of conformal field theories thus gives rise to two
immediate applications. Firstly one may try to gain an understanding
of stringy effects in quantum gravity by comparing the moduli space of
conformal field theories with the space of Einstein metrics for a
given class of backgrounds. Secondly one may assume that space-time is
in the form of flat four-dimensional Minkowski space times some
compact part $X$. The space of possible $X$'s leads to a space of
theories of particle physics (i.e., particle masses, couplings, etc.)
in four dimensional space time (see, for example, \cite{GSW:book}). In
this latter case $X$ has a Euclidean signature. Because of the
difficulty in analyzing conformal field theories associated to a
target space with indefinite signature we will need to restrict our
attention to the latter scenario. It should be expected however that
many of the features we observe in these lectures should carry over to
the former case of stringy quantum gravity of all of space-time.

In section \ref{s:CFT} we will deal with simple examples of
non-supersymmetric conformal field theories and their
moduli space to introduce the basic concepts we will require later in
these lectures. The basic example central to a great deal of work in this
subject will be that of $c=1$ theories and the linear sigma model
whose target space is a circle. The notion of \RoR\ duality appears
here and will be of some interest later in these lectures.

We will find that extending our ideas to more complicated examples is
very difficult to achieve in general. Because of this we are forced to
impose restrictions on the type of conformal field theories we
study. In particular we want to focus on conformal field theories
which are associated to some geometric target space (or perhaps some
slightly generalized notion thereof). We also impose
that the conformal field theory has $N$=2 supersymmetry. The
effect of this is to force the target space to be a space with a
complex structure. In terms of the flat four-dimensional Minkowski
space point of view these conditions amount the existence of a
space-time supersymmetry. For the purposes of these lectures we may
simply regard these conditions as providing us with enough structure
to use the tools of algebraic geometry.

In section \ref{s:torus} we will study the superconformal field theory
for a sigma model with a complex one-torus as the target space. This
will allow us to introduce the complex coordinates which prove to be
extremely useful for dealing with later examples.

Section \ref{s:K3t} will cover briefly the case of a K3 surface as
the target space.
In this case we have $N$=4 supersymmetry.
This section will also introduce the concept of a
``blow-up'' which is a key construction in algebraic geometry and thus
also appears naturally in the context of superconformal field
theories. This blow-up also appears to be of central importance to
understanding some global issues of the moduli space of $N$=2 theories
and so it will become something of a recurring theme in later sections.

In the sections discussed thus far we will find that using a metric as
an effective description of the target space suffices. For the rest of the
lectures however we will study examples which require more radical
approaches. In particular we will be required to think in terms of
algebraic geometry rather than differential geometry.

For the cases we discuss in the later sections, the moduli spaces
factor into two parts
$\cM(X)\cong\cM_A(X)\times\cM_B(X)$ (moduli some discrete symmetries
and so long as we are careful about the boundary points). In geometric
terms $\cM_A(X)$
corresponds to deformations of the (complexified) K\"ahler form on $X$
and $\cM_B(X)$ corresponds to deformations of the complex structure of
$X$.
The factor $\cM_B(X)$ turns out to be simple to understand and may be
analyzed classically.
In order to understand the structure of the moduli space of a
particular class of conformal field theories we will have to give
three interpretations to each point in $\cM_A(X)$:
\begin{enumerate}
\item
 The desired interpretation as a theory with some target space $X$ with
a specific K\"ahler form. This is the most difficult to analyze.
\item
 A theory with some flat target space containing $X$ with a specific
K\"ahler form. In some limit the fields in this theory are required to
live in $X$. This is the ``linear \sm'' of \cite{W:phase}.
\item
 A theory with some space $Y$, related to $X$ by ``mirror symmetry'',
where the point in moduli space specifies a complex structure on
$Y$.
\end{enumerate}
We will find that the third interpretation in terms of $Y$ provides the
simplest context in which to compute the moduli space but that we
require the linear \sm\ as an intermediary to translate between
interpretations on $X$ and $Y$ for each point in this space.

In section \ref{s:d=3} we will look at the simplest non-trivial
example of the above and explicitly compute $\cM_A(X)$. In section
\ref{s:phase} we will consider the more general case. Finally in section
\ref{s:conc} we present a few concluding remarks.


\section{Simple Models without Supersymmetry}	\label{s:CFT}

We will begin our discussion with the simplest \sm. For further
details and references as well as an excellent introduction to
conformal field theory the reader is referred to \cite{Gins:lect}.
Consider a field theory in a two-dimensional space
$\Sigma$ whose action is given by
\begin{equation}
  S = \frac{i}{8\pi\alpha^\prime}\int_\Sigma \partial x \bar\partial x
		\,d^2z,		\label{eq:bsm}
\end{equation}
where $x$ is real and we are using complex coordinate
$z=\sigma_1+i\sigma_2$ and its
conjugate $\bar z$ on the world-sheet, $\Sigma$.

Classically this action is conformally invariant and under
quantization one indeed obtains a conformal field theory with central
charge $c=1$. That is, the stress tensor $T(z)$ which in this case is
proportional to the normal ordered product $\mathord{:}\partial
x\partial x\mathord{:}(z)$
obeys the operator product expansion
\begin{equation}
  T(z)T(w) = \frac{c}{2(z-w)^4}+\frac2{(z-w)^2}T(w)
		+\frac1{z-w}\partial_wT(w)+\ldots	\label{eq:Vir}
\end{equation}
with $c=1$. In this way we say that we have a $c=1$ conformal field theory
description of the real line $\R$.

A simple generalization of (\ref{eq:bsm}) is to give the field $x$ an
index $i$ which runs $1,\ldots,n$. That is, the value of $x^i$ lies in
$\R^n$. We may also think of $x$ as defining a map $x\colon
\Sigma\to\R^n$. The stress tensor is now in the form
$\sum_i\normalord{\partial x^i\partial x^i}(z)$ and we
obtain (\ref{eq:Vir}) with $c=n$. Thus, at least in this simple case we
see that the central charge is a measure of the dimension of the
target space of the \sm.

Returning to the case $n=1$, we may make an alternative simple
generalization of our simple model by imposing periodic boundary
conditions on $x$. That is, we consider the target space to be a
circle of radius $R$ which implies $x \cong x+2\pi R$. Actually it is
convenient to rescale $x$ such that $x \cong x+2\pi$. Then the action
is
\begin{equation}
  S = \frac{iR^2}{8\pi\alpha^\prime}\int_\Sigma \partial x \bar\partial x
		\,d^2z.		\label{eq:bsmR}
\end{equation}
This is a conformal field theory with $c=1$ for any value of $R$
agreeing with the idea that a circle is a one-dimensional object
irrespective of its size! We have a family of field theories
parametrized by $R$.

The key purpose of these lectures is to discuss such families of
conformal field theories. The moduli space of theories is the space
mapped out by the parameters in the theory. Thus the moduli space for
theories on a circle appears at first sight to be the real half line
$\R_+$ mapped out by $R>0$. Certainly this is the moduli space for
circles. In order to be sure that we have the right moduli space
however it is important to ask the following question. Do two distinct
points in the supposed moduli space actually correspond to
identical conformal field theories? To answer this question we need to
know more about the theory than just the action (\ref{eq:bsmR}). While
different values of $R$ certainly give different actions it may be the
case that when we work out the spectrum of fields and their
correlators we end with identical field theories with different values
of $R$. Thus turns out frequently to be the case as we shall see.

Consider the local structure of the moduli space around a point
corresponding to the action $S_0$. For a field $\phi(z,\bar z)$ in
this theory we may build a new field theory
\begin{equation}
  S = S_0 + g\int_\Sigma \phi\,d^2z,
\end{equation}
for some small $g$. To maintain conformal invariance we require that
the term added be conformally invariant. This implies that $\phi$ must
be of weight (dimension) $(1,1)$ with respect to $z$ and $\bar z$.
Such an operator is called ``marginal''. It is important to realize
that the operator $\phi$ belongs to the field theory given by $S_0$
and not $S$. In order that we may ``transport'' this field to a new
one in the new theory which is also marginal we require that $\phi$
should not interact with itself to cause corrections to its own weight
\cite{Cardy:}.
Such a field is called ``truly marginal''. Such fields naturally span
the tangent space to any point in the moduli space.

It turns out that something rather special happens with
$R=\sqrt{\alpha^\prime}$ (see \cite{Gins:lect} for more details).
The resultant conformal field theory is completely
equivalent to the field theory that
corresponds to a string propagating on the group manifold $SU(2)$.
This has two striking consequence. Firstly $SU(2)$ is
three-dimensional and $c=1$ thus ruining our initial hope that the
central charge might give the dimension of the target space. Secondly,
the extra symmetries given by the affine algebra of $SU(2)$ map $\phi$
to $-\phi$ for this theory. Thus decreasing $R$ appears to be the same
as increasing $R$ away from this point in the moduli space.

Actually what we see here is the famous \RoR\ duality symmetry for a string on
a circle \cite{KY:rr,SS:rr}. It turns out that the partition function
of the string on a circle of radius $R$ is identical to that of
a string on a circle of radius $\alpha^\prime/R$. The reason for this
is quite simple to picture. As in normal quantum mechanics, the
momentum for the string going around the circle is quantized just like
a ``particle in a box''. This leads to ``momentum mode'' eigenstates
with energy proportional to $m^2\sqrt{\alpha^\prime}/R^2$ for $m\in\Z$.
Unlike the quantum theory of a particle we also have ``winding modes''
where the string wraps around the circle. Clearly the string must
wind an integer number, $n$, times round the circle forcing these modes to
be quantized too. The energy of these modes goes like $n^2 R^2/
(\alpha^\prime)^\frac32$. Thus, the \RoR\ symmetry appears as an
exchange in the r\^oles of $m$ and $n$, i.e., an exchange of winding
modes with momentum modes. (Note that in order to prove the equivalence
of two conformal field theories it is not sufficient just to show that
partition functions are identical. In this case however we know that
this duality holds near the point $R=\sqrt{\alpha^\prime}$ because of
the symmetry $\phi\leftrightarrow-\phi$ and we may integrate along
this marginal direction to extend the symmetry to a whole
$\Z_2$-symmetry acting on the line of $R$'s. This $\Z_2$ symmetry must
be the \RoR\ duality since this is the only symmetry of the partition
function with a fixed point at $R=\sqrt{\alpha^\prime}$.)

At this point our moduli space appears to be the half real line closed
at both ends by $R\geq\sqrt{\alpha^\prime}$ with the point at infinity
corresponding to $\R$, i.e., a circle of infinite radius. Actually
this is not the full moduli space of $c=1$ conformal field theories.
We may form orbifolds of theories we already have. To form an orbifold
we divide the theory out by a discrete symmetry $G$. This not only
projects out the non $G$-invariant states from the original theory but
also adds in ``twisted'' modes corresponding to open strings in the
original theory whose ends are identified under $G$. The circle
generically admits a $\Z_2$ symmetry by $x\mapsto-x$. This leads to a
new line of theories. We will have more to say about orbifolds later.
One may also divide the $SU(2)$ group manifold
at the $R=\sqrt{\alpha^\prime}$ point by any discrete subgroup of
$SU(2)$. Most of these subgroup lead to theories already accounted for
but the three exceptional subgroups (the bitetrahedral, bioctahedral
and biicosahedral) give new points \cite{DVV:c=1,Gins:c=1} in the
moduli space. Whether the full moduli space of unitary conformal field
theories with $c=1$ has now been accounted for is still to be proven.
However partial results towards this end have been reached
\cite{Kir:c=1}. The conjectured moduli space for $c=1$ theories is
shown in figure \ref{fig:c=1}. Interesting points have been labeled
--- the reader is again referred to \cite{Gins:lect} for more details.

\iffigs
\begin{figure}
  \centerline{\epsfxsize=9cm\epsfbox{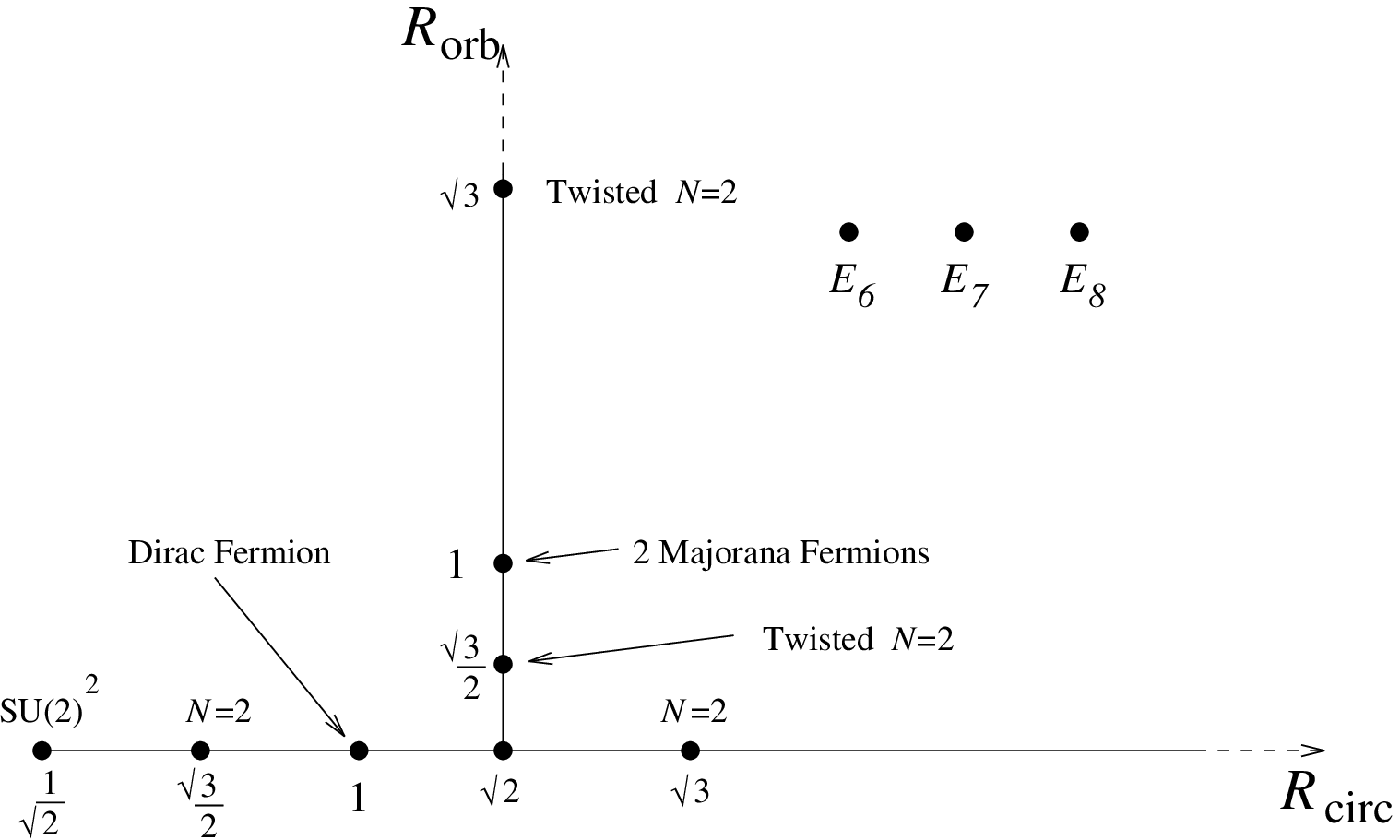}}
  \caption{The $c=1$ Moduli Space (for $\alpha^\prime=\ff12$).}
  \label{fig:c=1}
\end{figure}
\fi

The two generalizations of the real line we have considered thus far
were that of $\R^n$ and that of a circle. The next case is to combine
these to get a torus $(S^1)^n$. This will give us a conformal field
theory of central charge $c=n$. Again, we will not provide any
detailed proofs here but only state the results. Narain \cite{N:torus}
has shown that a string on an $n$-torus may be
specified by a lattice in a $2n$-dimensional space of indefinite
signature $(n,n)$. The generators of the lattice must have inner product in
the form of the matrix
\begin{equation}
\left(\begin{array}{ccccc}
  0&1&&&\\
  1&0&&&\\
  &&0&1&\\
  &&1&0&\\
  &&&&\ddots
\end{array}\right).	\label{eq:offH}
\end{equation}
Thus the question of the the moduli space of conformal field theories
is reduced to that of finding the moduli space of lattices in the
required form. The group $O(n,n)$ acts on any lattice to preserve
(\ref{eq:offH}) and this may be taken as the initial guess at the moduli
space. We need to identify points in the space $O(n,n)$ corresponding to
identical theories which requires us to take a quotient of this group.
Firstly we need to divide by $O(n)\times O(n)$. These rotations may be
viewed as rotations of the $n$-dimensional lattice defining the torus
itself (or the set of allowed winding modes) and the rotations of the
dual lattice (giving the set of allowed momentum modes). These
rotations have no effect on the underlying field theory. Thus our
candidate moduli space becomes the (right) coset $O(n,n)/((O(n)\times
O(n))$ (where the group $O(n)\times O(n)$ acts on $O(n,n)$ from the
right).

To complete the picture of the moduli space some more identifications
are required. These come from elements of $O(n,n)$ which act as
automorphisms of the lattice. It is easy to see that such
automorphisms must consist of $O(n,n;\Z)$, i.e., matrices in $O(n,n)$,
preserving (\ref{eq:offH}), which have purely integer entries.
Assuming we have accounted for all the necessary identifications our
moduli space can then be written as
\begin{equation}
  \cM_{\rm torus} \cong O(n,n;\Z)\backslash O(n,n)/(O(n)\times O(n)),
		\label{eq:mod-t}
\end{equation}
where the infinite discrete group $O(n,n;\Z)$ acts on $O(n,n)$ from
the left.

It is an instructive exercise to interpret this space from the point of
view of of the world-sheet action \cite{HNW:torus}. In the case of the
circle we could account for the one degree of freedom in the moduli
space as corresponding to the radius of the circle. The natural
generalization here would be to put a constant metric $g_{ij}$ on the
torus. Thus accounts for $\ff12n(n+1)$ dimensions but our moduli space
(\ref{eq:mod-t}) has $n^2$ dimensions. To account for these extra
terms one may add the term $i\varepsilon^{\alpha\beta} B_{ij}\partial_\alpha
x^i\partial_\beta x^j$ to the action, where $\alpha$ and $\beta$
indices label coordinates on the world-sheet and $B_{ij}$ is a
constant matrix which may be taken to be antisymmetric. Note the
factor of $i$ in this term
--- this is because of the transformation properties
of $\varepsilon^{\alpha\beta}$ under the Wick rotation from a
Minkowski signature world-sheet to the Euclidean signature we assume in these
lectures. Written in
terms of complex coordinates on the world-sheet our action then
becomes
\begin{equation}
  S = \frac i{8\pi\alpha^\prime}\int_\Sigma\left(g_{ij}-B_{ij}
	\right)\partial
	x^i\bar\partial x^j\,d^2z.	\label{eq:torsm}
\end{equation}

It turns out that in more general cases it is also natural to include
the antisymmetric $B$-term degree of freedom in the field theory.

In order to examine more complicated conformal field theory moduli
spaces it will be necessary to impose further structure. The structure
we will impose shall be supersymmetry. Thus, from this point on, all
conformal field theories considered will be superconformal field
theories. The \sm\ with target space given by a circle can be extended
in a straight-forward manner to a superconformal field theory. To do
so we add one free left-moving and one free right-moving Majorana-Weyl
fermion to our action. Such a free fermion is well-known to correspond
to the Ising model and contributes $c=\ff12$ to the central charge.
Our resulting theory will therefore have $c=\ff32$. The moduli space
of this theory is at first sight little more complicated than the
non-supersymmetric case. The only marginal operator in a generic theory is
the one that we already had acting to change the radius of the circle.
This should not be surprising since there are no deformations of the
free fermion part of the theory. The moduli space for these $c=\ff32$
theories contains figure \ref{fig:c=1} but it actually has more
branches. The reader is referred to \cite{DGH:c=3/2} where this
calculation was first performed for more details. The extra branches
in the moduli space arise because one may put different boundary
conditions on the fermions.

This moduli space of superconformal field theories with $c=\ff32$ is
certainly simpler than the space of all conformal field theories with
$c=\ff32$ so adding supersymmetry has simplified the problem.
Unfortunately this simplification is still not sufficient to study the
moduli spaces of non-trivial examples.

The main topic of these lectures will be conformal field theories with
$N=2$ world-sheet supersymmetry. We insist on $N$=2 superconformal
invariance in the both sectors and so such theories are often referred
to as $N$=(2,2) superconformal field theories.
It is this extended supersymmetry
structure which allows for a dramatic simplification of the problem.
The fundamental reason for this is that in this case one works
exclusively with objects taking values in the complex numbers $\C$
rather than $\R$. Holomorphicity then allows many problems to be
solved.


\section{The Complex Torus}   	\label{s:torus}

\subsection{The Analytic Approach}

To begin our discussion of $N$=2 theories let us try to build the
simplest theory with a non-trivial moduli space in analogy with the
$c=1$ theories and the $N=1, c=\ff32$ theories above. Adding a second
free moving fermion to each sector forces us to add another free boson
to complete the $N=2$ supermultiplet. It appears that our simplest
model will have $c=3$. Actually the moduli space of such theories
turns out to be very messy. Extra branches like those that appeared in the
$c=\ff32$ theories of \cite{DGH:c=3/2} proliferate in this case.
To bring this situation under control let us impose a further
condition on our theories.

Each field in our $N$=2 superconformal field theory has charges
$(q,\bar q)$ under the left-moving and right-moving $U(1)$ currents
implicit in the $N$=2 algebra. We will insist that all NS fields in
our theory will have $q,\bar q\in\Z$. Such a constraint appears
naturally in string theory\footnote{Actually the constraint imposed in
superstring theory is that these charges be {\em odd\/} integer. For
the purposes of these lectures however we will not need this stronger
condition.} --- it is required for ``spectral flow'' to
be a symmetry for fields appearing in the partition function.
The reader is referred to \cite{BRG:lect} for more details.

Let us compute the complete moduli space for such conformal field
theories with $c=3$. The following argument was first presented in
\cite{Gep:torus} and we present here only an outline.
For the time being we will assume $\alpha^\prime=\ff12$ but this will
change later.
Using the usual
notation $T(z)$, $G^\pm(z)$ and $J(z)$ for the fields generating the
$N$=2 algebra in the left sector, we have
\begin{equation}
  J(z)J(w)=\frac1{(z-w)^2}+\ldots
\end{equation}
This implies there exists some free boson $\phi$ such that
\begin{equation}
  J(z) = i\partial\phi(z).
\end{equation}
Furthermore, we can define fields, $\hat G^\pm$, which are neutral
under this $U(1)$ current:
\begin{equation}
  G^\pm = \hat G^\pm\,\normalord{e^{i\phi}}.
\end{equation}
It follows that
\begin{equation}
  \hat G^+(z)\hat G^-(w) = \frac2{(z-w)^2}+\ldots
\end{equation}
This in turn implies the existence of a complex free boson $H$ such
that
\begin{equation}
  \hat G^+(z)=\sqrt{2}\partial H,\quad \hat G^-(z)=\sqrt{2}\partial
  H^\dagger.
\end{equation}

The free boson $\phi$ by itself forms a $c=1$ conformal field theory
as we described earlier. Precisely which conformal field it
corresponds to is given by the periodicity of the field $\phi$ --- i.e., the
radius of the circle on which $\phi$ lives. This is fixed by our
quantization condition on the fields in our theory. A field of
charge $q$ will contain a $\normalord{e^{iq\phi}}$ factor. The
existence of charge one fields implies that the radius is 1. From
figure \ref{fig:c=1} we see that this corresponds to a Dirac fermion.
That is, we have two fermions with matching boundary conditions. Call
this a complex fermion $\psi$. We then have that
\begin{equation}
  J(z) = \normalord{\psi^\dagger\psi}(z).
\end{equation}

All our degrees of freedom in the moduli space appear to be
encapsulated in the field $H$. The first possibility to consider is
that this corresponds to some two-dimensional torus. The moduli space
of such $c=2$ conformal field theories was studied in
\cite{DVV:torus}. Actually we already know from we what have said that
it should be
\begin{equation}
  O(2,2;\Z)\backslash O(2,2)/(O(2)\times O(2)).
\end{equation}
Neglecting factors of $\Z_2$, one knows that
\begin{equation}
  O(2,2)/(O(2)\times O(2)) \cong \left[Sl(2)/U(1)\right]^2.
\end{equation}
We may map $Sl(2)/U(1)$ into the upper half complex plane by
\begin{equation}
\left(\begin{array}{cc}a&b\\c&d\end{array}\right)\mapsto
	\frac{ai+b}{ci+d},\quad\hbox{where\ }
  \left(\begin{array}{cc}a&b\\c&d\end{array}\right) \in Sl(2).
\end{equation}
Thus, modulo issues of discrete group actions our moduli space is two
copies of the upper-half complex plane. This space is referred to as
the ``Teichm\"uller space''. The actual moduli space is the
Teichm\"uller space divided by some discrete group action. The group
$O(2,2;\Z)$ contains $Sl(2,\Z)\times Sl(2,\Z)$. These
$Sl(2,\Z)$ groups act on the two upper half planes by
\begin{equation}
\left(\begin{array}{cc}a&b\\c&d\end{array}\right)\colon z\mapsto
	\frac{az+b}{cz+d},\quad\hbox{where\ }
  \left(\begin{array}{cc}a&b\\c&d\end{array}\right) \in Sl(2,\Z).
		\label{eq:modgroup}
\end{equation}
Thus in our moduli space each upper half plane is divided by this
modular group. To picture the moduli space we should therefore
consider one fundamental domain for this group for each upper-half plane.

In our haste in building the moduli space thus far we have neglected a
few $\Z_2$ factors. The correct inclusion of these divides the space
by two further $\Z_2$-actions (see \cite{GMR:Z2} for this
calculation). Let us label the two half planes by
$\sigma,\tau\in \C; \Img(\sigma),\Img(\tau)>0$. As well as the modular
group action (\ref{eq:modgroup}) on each of these parameters we also
have
\begin{equation}
  \eqalign{
  \mu\colon (\sigma,\tau)&\mapsto (\tau,\sigma),\cr
  \xi\colon (\sigma,\tau)&\mapsto (-\bar\sigma,-\bar\tau).\cr}
\end{equation}
The resulting moduli space is shown in figure \ref{fig:mod-t}.

\iffigs
\begin{figure}[t]
  \centerline{\epsfxsize=11cm\epsfbox{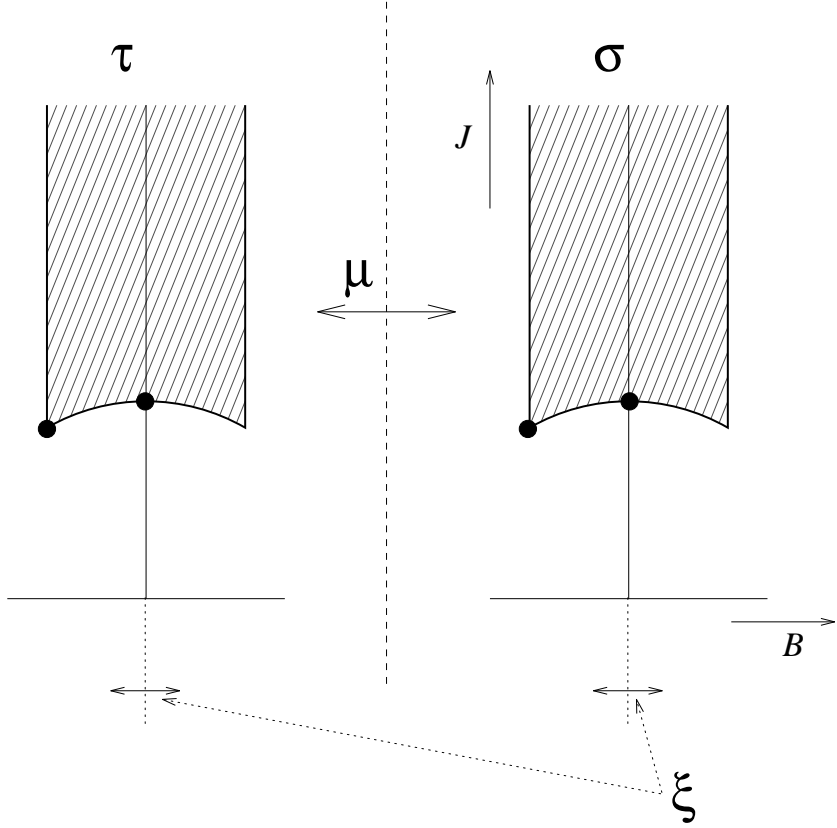}}
  \caption{The $d=1$ Moduli Space.}
  \label{fig:mod-t}
\end{figure}
\fi

When the moduli space of $c=2$ theories was investigated in
\cite{DVV:torus} it was noted that figure \ref{fig:mod-t} was not the
complete moduli space. In general one could take orbifolds of the
2-torus to obtain other conformal field theories with $c=2$. In the
context of our $c=3$ superconformal field theory any
action of a group on this torus which is not simply a translation
would act non-trivially on $G^\pm(z)$. Thus dividing by such a group
would ruin the $N$=2 superconformal invariance. One concludes
therefore that the moduli space as shown in figure \ref{fig:mod-t}
includes all valid $N=2,c=3$ superconformal field theories.

It will prove extremely useful to analyze figure \ref{fig:mod-t} from
the point of view of complex geometry. Let us first note that a
2-torus is a complex manifold of one dimension. Now rewrite the
action (\ref{eq:torsm}) in terms of complex coordinates $x^i$ on the
target space:
\begin{equation}
  S = \frac i{4\pi\alpha^\prime}\int\left\{g_{i\bar\jmath}
        (\partial x^i\bar\partial x^{\bar\jmath}+
        \bar\partial x^i\partial x^{\bar\jmath})
        -iB_{i\bar\jmath}(\partial x^i\bar\partial x^{\bar\jmath}-
        \bar\partial x^i\partial x^{\bar\jmath})
        \right\}\,d^2z.         \label{eq:sm}
\end{equation}

In terms of the action (\ref{eq:sm}) we appear to have the following
degrees of freedom: complex structure on $X$ and a choice of
$g_{i\bar\jmath}$ and $B_{i\bar\jmath}$. Can we interpret the moduli
space of figure \ref{fig:mod-t} in terms of these? The complex
structure part is easy. It is well-known that the moduli space of a
torus is given by the upper half plane divided by $Sl(2,\Z)$. We may
thus take $\tau$ to represent the complex structure in the usual way,
i.e., the complex structure is that of a torus constructed from the
complex plane by dividing by the translations given by 1 and $\tau$
where $\Img(\tau)>0$.

It may be shown \cite{DVV:torus} that the complex parameter $\sigma$ may
be built from the remaining degrees of freedom:
\begin{equation}
  \sigma = \frac1{4\pi^2\alpha^\prime}(B_1+iJ_1).
\end{equation}
The real numbers $B_1$ and $J_1$ describe the degrees of freedom of
the metric and the $B$-term as follows. Introduce a constant 2-form $e$ such
that $\int_X e=1$. Thus $e$ generates $H^2(X,\Z)$. Now write
\begin{equation}
  \eqalign{B=B_1e,\cr J=J_1e,\cr}
\end{equation}
where
\begin{equation}
  \eqalign{B=\ff i2 B_{i\bar\jmath}\,dx^i\wedge dx^{\bar\jmath},\cr
	J=\ff i2 g_{i\bar\jmath}\,dx^i\wedge dx^{\bar\jmath}.\cr}
		\label{eq:BJdef}
\end{equation}
 From now on, to
avoid cumbersome factors, let us set units so that
$4\pi^2\alpha^\prime=1$. The element in $Sl(2,\Z)$ which takes
$\sigma\to\sigma+1$ is easy to
explain as follows. We may rewrite the
action (\ref{eq:sm}) as
\begin{equation}
  S = S_g + 2\pi i\int_\Sigma x^*(B),
\end{equation}
where $S_g$ does not depend on $B$ and $x^*(B)$ denotes the pull-back
of the two-form, $B$,  from $X$ to the
world-sheet $\Sigma$ via the map $x\colon\Sigma\to X$.
First note that this term depending on $B$ is
``topological'' --- it depends only on the cohomology class of $B$.
Thus deformations of $B$ do not affect $S$. This means that the
equations of motion do not depend on $B$. At the quantum level
$B$ is important but correlation functions can only ``see'' $B$
via the expression $\exp(-S)$ in the path integral. Thus, if we shift
$B$ by an element of $H^2(X,\Z)$ then $S$ shifts by $2\pi in$ for
$n\in\Z$ and thus the correlation functions are invariant. This will
be a general symmetry of any theory based on the action (\ref{eq:sm}).
In the case of the torus it corresponds to $\sigma\to\sigma+1$.

The other generator of $Sl(2,\Z)$ not yet accounted for takes $\sigma\to
-1/\sigma$. If we take $B$ to be zero, this amounts to inverting
$J_1$. Since $J_1$ corresponds to the area of the
torus, this action amounts to the \RoR\ duality acting on the torus.
The $\Z_2$-action $\xi$ also has a straight-forward explanation. One
can show that it amounts to taking the complex conjugate of the field
theory. In target space terms this amounts to taking the conjugate
complex structure on the target space and changing the sign of $B$.

The last symmetry, generated by $\mu$, has no classical explanation.
It is ``mirror symmetry'' (see \cite{BRG:lect} for
an account of the evolution of this phenomenon).
For some points in this moduli space we may also view this as coming
from \RoR\ duality. This occurs as follows. Let the torus be generated
by dividing the complex plane by translations by $2\pi R_1$ and $2\pi R_2$ in
orthogonal directions and let $B$ be zero. This gives
\begin{equation}
  \eqalign{\tau&=i\frac{R_1}{R_2}\cr
  \sigma&=\frac{i}{\alpha^\prime}R_1R_2.\cr}
\end{equation}
The \RoR\ duality acting on $R_2$ gives the mirror map $\mu$. It is
important to realize however that this is a very special case we will
not expect there to be any relationship between mirror symmetry and
\RoR\ dualities in general.

Note that prior knowledge of the
existence of mirror symmetry would have been a powerful tool in
building this moduli space. Having built the Teichm\"uller space we
could have generated the whole modular group from the following:
\begin{enumerate}
  \item The modular group for the classical moduli space of complex
structures, i.e., $Sl(2,\Z)$.
  \item The $\Z_2$ generated by the mirror map.
  \item The $\Z_2$ generated by complex conjugation plus change in
sign of $B$.
\end{enumerate}

Now let us discuss some of the philosophy behind this latter approach
to building the modular group and thus moduli space and contrast it to
the method that we used formerly in this section.
The first thing to note is that this latter approach
differs markedly from the approach one might consider more natural in
physics. As physicists we are used to analyzing space-time in terms of
a Riemannian metric. Thus when a theory is presented in the form
(\ref{eq:sm}) and one is asked the question ``What is the moduli space
for such theories?'', one's first reaction will be to find the moduli
space of allowed metrics $g_{i\bar\jmath}$. In the cases presented thus
far in these lectures, the metric is constant and the theory can be
solved. Indeed this is the method by which we effectively found the
moduli space. As we shall see, when the metric is not so trivial, life
becomes considerably more difficult. The stunning feature of the
program for building the moduli space using mirror symmetry as
outlined in the three steps above is that nowhere did we mention the
metric! One might at first think that the complex structure encodes at
least part of the metric data since one may think of it as being used
to go from the real metric to the Hermitian metric used in
(\ref{eq:sm}). Actually this need not be the case as we now explain.

The branch of geometry usually used by physicists is differential
geometry. This is implicit when one uses Riemannian geometry. Given a
manifold $X$, one looks very closely at a very small part of it and
assumes it looks roughly like flat space. By gluing many such small
parts together and defining connections one can build a global
picture. There is another method however --- namely {\em algebraic\/}
geometry. In algebraic geometry (over $\C$) one considers recovering
information concerning the geometry of a space by considering
holomorphic functions on that space. Equivalently one may build a
space as the zero locus of some function(s) defined on a simpler space
(such as a complex projective space). One need never mention the
metric when studying algebraic geometry. Indeed, one need not insist
that the space in question is a manifold and thus it may be difficult
to define a metric on our space anyway. The important
point to notice is that the moduli space of complex structures of a
space may be determined using the methods of algebraic geometry.

The situation in $N$=2 theories is actually twice as algebraic as
classical geometry thanks to mirror symmetry. In most cases (but not
always as we see in section \ref{s:K3t}) the moduli space roughly
factorizes into two parts. One part describes the complex structure
and the other half gives the K\"ahler form (i.e., Hermitian metric)
and $B$. Classically the first half can be determined by
algebraic geometry whereas the second half is not such a naturally algebraic
object. In string theory
the mirror map relates the second half to a complex structure
calculation allowing both halves to be calculated using algebraic
geometry! Certainly within the context of $N$=2 theories it would seem
unfortunate that physicists have grown to be biased in favour of
differential rather than algebraic geometry for analyzing problems. In
fact it is the case that not only is algebraic geometry the more
convenient setting for analyzing such string theories but that
differential geometry can lead to incorrect conclusions as we shall
see later. This should not
really be a surprise. The fundamental assumption in differential
geometry is that if we look closely enough at a small piece of space
then it looks like flat space. Why should this be the case in string
theory?

\subsection{The Algebraic Approach}	\label{ss:t-alg}

Let us now reanalyze the moduli space of the torus from an algebraic
point of view. This may appear to be unnecessarily cumbersome compared
to the analytical approach used earlier in this section. However,
since we make no reference to metrics but rather use mirror symmetry
this method will be easier to generalize to more complicated examples.

Consider the complex projective space $\P^2$ with homogeneous
coordinates $[x_0,x_1,x_2]$. Define the hypersurface $X$ by the
equation
\begin{equation}
  f=x_0^3+x_1^3+x_2^3-3\psi x_0x_1x_2=0.	\label{eq:ell}
\end{equation}
An object with one complex dimension defined algebraically is called an
``algebraic curve''. Algebraic geometers tend to name things as if
complex dimensions were real dimensions --- thus the term ``surface''
refers to two complex dimensions. The term ``algebraic variety'' is used
for the general case of any dimension. An algebraic curve is a Riemann
surface and so the topology is defined by the genus. A curve of genus
zero is called ``rational'' and a curve of genus one is called
``elliptic''. A straight-forward Euler characteristic calculation
(see, for example, \cite{Hub:book})
shows that (\ref{eq:ell}) defines an elliptic curve. Thus we have an
algebraic description of our torus.

As we vary $\psi$ in (\ref{eq:ell}) we vary the complex structure of
$X$. In general there is a rather subtle relationship between deformations
of the polynomial(s) defining the algebraic variety and deformations
of complex structure. See \cite{GH:poly} for a thorough treatment of
this question. In many simple cases however each nontrivial deformation of
the defining polynomial (i.e., deformations which cannot be undone by a
linear redefinition of the coordinates) define a deformation of
complex structure and all deformations of complex structure are
obtained this way. This is the case here for the elliptic curve. Note
that linear changes of coordinates $(x_0,x_1,x_2)\to(x_0+x_1+x_2,x_0+\omega
x_1+\omega^2 x_2,x_0+\omega^2 x_1+\omega x_2)$ and
$(x_0,x_1,x_2)\to(\omega x_0,x_1,x_2)$,
where $\omega$ is a on-trivial cube root of unity, induce the following
transformations of $\psi$:
\begin{equation}
  \eqalign{
  \psi &\to -\frac{2+\psi}{1-\psi}\cr
  \psi &\to \omega\psi.\cr}
	\label{eq:ellmod}
\end{equation}
The moduli space for complex structures of an elliptic curve may be
parametrized by $\psi$ modulo the transformations (\ref{eq:ellmod}).

Note that something special happens at $\psi=1$. In this case $f$
factorizes:
\begin{equation}
  f = (x_0+x_1+x_2)(x_0+\omega x_1+\omega^2 x_2)(x_0+\omega^2 x_1+\omega x_2).
\end{equation}
Thus $f=0$ has three components given by the vanishing of each factor.
Each factor is a hypersurface defined by a linear equation in $\P^2$
which is a rational curve. Thus $X$ consists of three rational curves.
Each pair of rational curves intersect at one point making a total of
three points of intersection. This space clearly is not a manifold. In
general singularities will occur along a hypersurface defined by $f$
when there is a solution to the equations
\begin{equation}
  f=\frac{\partial f}{\partial x_0}=\frac{\partial f}{\partial x_1}=
	\ldots=0.	\label{eq:smooth}
\end{equation}
In this case we have solutions when $\psi=1,\omega,\omega^2,\infty$.
These four solutions are mapped to each other by (\ref{eq:ellmod}).
This value of $\psi$ is not really allowed in the moduli space for the
elliptic curve since it does not give a smooth elliptic curve. For
many purposes however it is useful to add it in to form a
compactified moduli space.

Let us now relate $\psi$ to the other parameterization of the
complex structure using $\tau$ from the previous section. Consider two
one-cycles $\gamma_0,\gamma_1$ on $X$ that generate $H_1(X)$. If we
cut $X$ along these cycles we obtain a parallelogram that can be put
on the complex plane. Let us use $\xi$ to denote the complex number
parametrized by this plane. If we rescale by a complex number so that one of
the edges lies along the line from 0 to 1, the other edge will lie
along 0 to $\tau$. We may also insist that $\Img\tau>0$ (since if this
fails, simply exchange the cycles).
Thus, if we define $\Omega=f\,d\xi$ as a holomorphic
1-form on $X$ where $f$ is a constant we have
\begin{equation}
  \tau= \frac{\displaystyle\int_{\gamma_l}\Omega}
        {\displaystyle\int_{\gamma_0}\Omega}. \label{eq:period}
\end{equation}
That is, $\tau$ is defined by the ratio of two ``periods''. It turns
out that such periods satisfy a differential equation known as the
Picard-Fuchs equation. In general it is rather cumbersome to set up
the machinery for analyzing these differential equations and solutions
so here we present only a quick outline of one of the many methods of
derivation of the periods.

First let us find a representative for $\Omega$. In an affine patch
where we put $x_0=1$ and use $x_1$ and $x_2$ as affine coordinates we let
\begin{equation}
  \Omega=\left(\frac{\partial f}{\partial x_2}\right)^{-1}\,dx_1.
\end{equation}
This is clearly a holomorphic 1-form. One may also show that it is
everywhere finite by going to the other patches. Thus it may be used
to represent $\Omega$. We now follow \cite{lots:per} in finding the periods.
Consider the 1-cycle $\gamma_0$ in the
elliptic curve defined by $x_0=1$, $|x_1|=\epsilon_1$ and $x_2$ by
the unique solution to (\ref{eq:ell}) such that $x_2\to0$ as
$\psi\to\infty$. Here, $\epsilon_1$ is a small positive real
number. This 1-cycle is enclosed by the 2-cycle $\Gamma_0$ defined by
$x_0=1, |x_1|=\epsilon_1, |x_2|=\epsilon_2$ defined in the ambient
$\P^2$. We may thus use Cauchy's theorem to give
\begin{equation}
  \eqalign{\varpi_0&=\int_{\gamma_0}\Omega\cr
   	&=\int_{\gamma_0} \left(\frac{\partial f}{\partial x_2}\right)^{-1}
            \,dx_1\cr
	&=\int_{\Gamma_0} \frac{dx_1\,dx_2}{f}\cr
	&=\int_{\Gamma^\prime_0} \frac{dx_0\,dx_1\,dx_2}{f}.\cr}
		\label{eq:pers}
\end{equation}
We have neglected overall constant factors at each stage since these
are irrelevant. The last step in (\ref{eq:pers}) is a somewhat formal
manipulation to make things more symmetric. $\Gamma^\prime_0$ is the
3-cycle in $\C^3$ defined by $|x_i|=\epsilon_i$.

Let us now rewrite the defining equation in a more general form which
allows us to lift questions into $\C^3$:
\begin{equation}
  \varpi_0=\int_{\Gamma^\prime_0} \frac{dx_0\,dx_1\,dx_2}
	{a_0x_0^3+a_1x_1^3+a_2x_2^3+a_3x_0x_1x_2}.
\end{equation}
It is now a simple matter to show that
\begin{equation}
  \frac{\partial^3}{\partial a_0\partial a_1\partial a_2}\varpi_0
	=\frac{\partial^3}{\partial a_3^3}\varpi_0.
\end{equation}
Actually this last differential equation did not depend on our choice
of cycle and will be satisfied by any of the cycles. Thus all periods
$\varpi$ may be obtained this way.

Careful analysis \cite{Bat:var} of the relationship between this
affine point of view and our desired interpretation in the projective
space shows that $\varpi=a_3^{-1}f(z)$, where
$z=-27a_0a_1a_2/a_3^3=\psi^{-3}$ for some function $f$.
For any period on our elliptic curve we thus obtain
\begin{equation}
  z\frac d{dz}\left\{\left(z\frac d{dz}\right)^2f-z\left(z\frac
  d{dz}+\ff13\right)\left(z\frac d{dz}+\ff23\right)\right\}f=0.
	\label{eq:bigPF}
\end{equation}
This equation will have three independent solutions although we expect
only two, since we only have two linearly independent 1-cycles. The
extra solution arises because we did the analysis in $\C^3$. The extra
solution may be removed by omitting the initial $z\ff d{dz}$ in
(\ref{eq:bigPF}). (This may be shown by analyzing the monodromy of the
extra solution.) The reader is referred to \cite{Mor:PF} for an
alternative derivation.
The resulting second order ODE is the Picard-Fuchs equation and is a
hypergeometric differential
equation. A general solution is
of the form
$f=Af_A(z)+Bf_B(z)$ where
\begin{equation}
  \eqalign{
    f_A(z)&=1+\ff29z+\ff{10}{81}z^2+\ff{560}{6561}z^3+\ldots\cr
    f_B(z)&=f_A(z).\log z+\ff59z+\ff{19}{54}z^2+\ldots\cr}
		\label{eq:fAB}
\end{equation}

The value $z=0$ corresponds to our singular elliptic curve and so to
map to the fundamental domain as shown in figure \ref{fig:mod-t} we
map this to $\tau=i\infty$. This point may be considered the fixed
point of the $\tau\to\tau+1$ symmetry. To recover this symmetry when
expressing $\tau$ as a ratio of periods one is forced to choose
\begin{equation}
  \eqalign{
  \tau&=\frac1{2\pi i}\left(\frac{f_B(z)}{f_A(z)}+k\right)\cr
  &=\frac1{2\pi i}\left(\log z+k+\ff59z+\ff{37}{162}z^2
	+\ff{2669}{19683}z^3+\ldots\right),\cr}
\end{equation}
for some constant $k$.

In order to determine $k$ we need to be able to identify another point in
our moduli space. For $\psi=0$ the elliptic curve admits a $\Z_3$
symmetry generated by $(x_0,x_1,x_2)\mapsto(\omega x_0,
x_1,x_2)$. This action has three fixed points:
$[0,-1,1],[0,-\omega,1],[0,-\omega^2,1]$. The only torus that admits a $\Z_3$
symmetry with fixed points occurs for $\tau=\omega$. In this case the
symmetry is generated by $\xi\mapsto \omega\xi$. Therefore we see that
$\psi=0$ is equivalent to $\tau=\omega$. Unfortunately the series
(\ref{eq:fAB}) fail to converge for $|z|>1$. In order to extend to
this region we need to analytically continue these functions. This is
done by representing the functions as Barnes integrals:
\begin{equation}
  \eqalign{
    f_A&=\frac1{2\pi i}\int_C \frac{\Gamma(3s+1)\Gamma(-s)}
	{\Gamma^2(s+1)}\left(-\frac z{27}\right)^s\,ds\cr
    f_B&=\frac1{2\pi i}\int_C \frac{\Gamma(3s+1)\Gamma^2(-s)}
	{\Gamma(s+1)}\left(\frac z{27}\right)^s\,ds+f_A.\log27,\cr}
\end{equation}
where $C$ is the contour running from $-\epsilon-i\infty$ to
$-\epsilon+i\infty$ along $\Rea(s)=-\epsilon$ for some positive real
number $\epsilon$. Closing the contour to the right recovers
(\ref{eq:fAB}). Enclosing to the left recovers different series valid
for $|z|>1$ which are the analytic continuations of (\ref{eq:fAB}).
Taking the limit $z\to\infty$ determines $k=-\log27$.

To complete our moduli space for the $N$=2 theories we copy the above
structure. We introduce a new algebraic parameter $y$ and set
\begin{equation}
  B_1+iJ_1=\sigma=\frac1{2\pi i}\left(\log\frac y{27}+k+\ff59y+\ff{37}{162}y^2
	+\ff{2669}{19683}y^3+\ldots\right),
\end{equation}
as the mirror of our complex parameter $\tau$. Mirror symmetry now
exchanges $y$ and $z$ and hence $\sigma$ and $\tau$. Simultaneous
complex conjugation of $y$ and $z$ sends $\tau\to\bar\tau$ and
$B\to-B$ as desired.

We have thus built a complete description of the moduli space
without requiring a metric on the target space.


\section{The K3 Surface}	\label{s:K3t}

Let us consider a target space that is not flat. That is, we allow
$g_{i\bar\jmath}$ and $B_{i\bar\jmath}$ in (\ref{eq:sm}) to vary over
$X$. It is not possible, in general, to solve this model
exactly. This model is known as the non-linear \sm. One way of
analyzing this model is to look at the $\beta$-functions in
perturbation theory \cite{Cal:sm}. For conformal invariance these beta
functions must vanish. Suppose $R$ is some characteristic radius of
the space $X$ is some vague sense. One will expect the perturbation
theory to be an expansion roughly in $\alpha^\prime/R^2$. Thus this
method should be reliable in the ``large radius'' limit.

A simple solution to the vanishing of the $\beta$-functions at one
loop is given by \cite{Cal:sm}
\begin{equation}
  \eqalign{dB&=0\cr R_{i\jmath}&=0,\cr}		\label{eq:CY}
\end{equation}
and we demand that the metric $g_{i\bar\jmath}$ be K\"ahler.
That is,
it may be written in terms of a closed (1,1)-form $J$ given by
(\ref{eq:BJdef}). A manifold
which admits a Ricci-flat K\"ahler metric is called a ``\CY\
manifold''.\footnote{In these lectures we will often allow
degenerations of such a manifold. We therefore often use the term
``space'' rather than ``manifold''.}
The first
condition $dB=0$ may be thought of as an assumption from which the
\CY\ conditions follow. As an alternative one might introduce a
non-trivial field $H=dB$, indeed one may allow $H$ to be a
cohomologically non-trivial 3-form so that $B$ is only locally
defined. WZW models on a group manifold \cite{W:WZW} are an example of this.
We will not concern ourselves with such models in these
lectures mainly because they are very difficult to analyze for
anything but the simplest metric. One might
also hope that any such model is equivalent to some \CY\ model (if the
class of \CY\ spaces is generalized in some way).

Therefore, for the purposes of these lectures, if an $N$=2 theory has
any large radius interpretation then it must be a \CY\ space.
The constraint of Ricci-flatness is actually topological in nature.
The first Chern class of the tangent bundle on a manifold, which is an
element of $H^2(X,\Z)$, is defined
in terms of the Ricci curvature and so must be trivial for a \CY\
space. In these lectures we will denote this condition by
$K=0$.\footnote{This notation comes from algebraic geometry where $K$
denotes the canonical divisor. This divisor is associated to the first Chern
class of the canonical bundle which is turn is equal to the first
Chern class of the cotangent bundle.}
Actually, due to a theorem of Yau \cite{Yau:}, this argument works in
reverse. That is, given a manifold with $K=0$ one may prove the
existence of a Ricci-flat K\"ahler metric. To be more specific, given
a \CY\ manifold (with a given complex structure) and a suitable cohomology
class in $H^{1,1}(X)$ there is a {\em
unique\/} Ricci-flat metric such that the K\"ahler form is a
representative of the cohomology class. Thus rather than concerning
ourselves with details of the metric one can specify the cohomology
class of $J$ and let Yau's theorem handle the rest.

On the face of it, the non-linear \sm\ with target space $X$, has the
following degrees of freedom.
\begin{enumerate}
\item The complex structure of $X$.
\item The cohomology class of the K\"ahler form, $J$.
\item The cohomology class of $B$ modulo $H^2(X,\Z)$.
\end{enumerate}

Let us
use $d$ to denote the complex dimension of the target space. For $d=1$
the \CY\ space is the torus and we studied that in the last
section. For $d=2$ there are two possibilities. Firstly there is the
two complex dimensional torus --- again a flat space. Secondly one
might have a ``K3 surface''. Before moving on to $d=3$ we will briefly
analyze some aspects of this K3 case. Although this is not entirely
relevant for the case $d=3$ it will allow us to introduce \CY\ orbifolds
which will be of some importance later on.

In the case $d=1$ we found that the superconformal algebra provided a
very restricting set of conditions for the theory. It was this that
forced the target space to be a torus. For $d\geq2$ we lose some of the
power of this method but for $d=2$ one may still impose certain
restrictions. In this case we can define the boson $\phi(z)$ by
\begin{equation}
  J(z)=i\sqrt{2}\partial\phi(z).
\end{equation}
We also have the fields
\begin{equation}
  \Omega^\pm(z) = \normalord{e^{\pm i\sqrt{2}\phi}}(z),
\end{equation}
which are present in any non-trivial theory with NS fields with
integer $U(1)$ charges. The fields $J(z),\Omega^\pm(z)$ together form
an $SU(2)$ affine algebra \cite{Eg:K3}. Thus the $U(1)$ of the $N$=2
superconformal algebra has been elevated to $SU(2)$. In fact, the two
fields $G^\pm(z)$ may be split up into 4 fields transforming as a ${\bf
2}+{\bf 2}$ representation of $SU(2)$. In this way, the algebra is
extended to an $N$=4 superconformal algebra. That is to say, in the
case $d=2$, any $N$=2 superconformal field theory with NS fields with
integer $U(1)$ charges is automatically an $N$=4 superconformal field
theory.

This $N$=4 supersymmetry has a striking effect on the perturbation
theory for the non-linear \sm. One may show
\cite{AGG:N=4,Hull:N=4,BS:N=4} that there are no corrections to
(\ref{eq:CY}) at any loop order, nor indeed any nonperturbative
corrections. One may also show that the target space has a
quaternionic structure as well as a complex structure.

Now let us turn our attention to K3 itself. What exactly is a K3
surface? One way of building one is in the form of an orbifold.
Consider the two complex dimensional torus constructed by taking a
quotient of $\C^2$, with coordinates $(\xi_1,\xi_2)$, by the
translations $(1,0),(i,0),(0,1),(0,i)$. In other words we take the
product of two tori with $\tau=i$ as defined in the last section. The
group $\Z_2$ generated by $g\colon
(\xi_1,\xi_2)\mapsto(-\xi_1,-\xi_2)$ fixes the 16 points
$(0,0), (\ff12,0), (\ff12i,0), (\ff12+\ff12i,0), (0,\ff12),\ldots
,(\ff12+\ff12i,\ff12+\ff12i)$. This means that when we build the space
that is the quotient of the complex two-torus by this $\Z_2$ we will
have a space with 16 singularities. Each of the 16 singularities looks
locally like $\C^2/\Z_2$ where the $\Z_2$ in the denominator is
generated by $(\xi_1,\xi_2)\mapsto(-\xi_1,-\xi_2)$. This latter space
has an isolated singularity at the origin.

A space with quotient singularities is known as an orbifold.
Because such spaces are not manifolds, na{\"\i}ve application of
differential geometry would be inappropriate.
They are simple to deal with in terms of
algebraic geometry however. To see how such objects appear in our
moduli spaces we need to introduce the concept of a ``blow-up''.

To begin with let us define the space
\begin{equation}
  \O(-1)=\{[a,b],(x,y)\in\P^1\times \C^2;\,ay=bx\}.
\end{equation}
Clearly this is a two-dimensional space (there is one constraint in a
three dimensional space). If $(x,y)\neq(0,0)$ then $a/b$ is determined
and we fix a point on $\P^1$. Thus, away from the origin of $\C^2$
there is a one-to-one map between $\C^2$ and $\O(-1)$. At the origin
of $\C^2$ there is no constraint on $[a,b]$ so we recover the whole
$\P^1$. Thus $\O(-1)$ looks like $\C^2$ with the origin removed and
replaced by $\P^1$. One may also check that $\O(-1)$ is smooth from
(\ref{eq:smooth}). The space $\O(-1)$ is said to be obtained from
$\C^2$ by ``blowing-up'' the origin. Given a space, one may produce an
infinite set of spaces by blowing-up smooth points. This might at
first appear an alarming prospect from the point of view of trying to
classify \CY\ spaces but it turns out that if we take a \CY\ space and
blow-up a smooth point then the resulting space has $-K>0$. That is,
the resulting
space will not be \CY\ (although it is still K\"ahler). The usefulness of the
blowing-up construction in the context of \CY\ spaces stems from what
happens when we try to blow-up singular points as we now see.

Let us return the the space $\C^2/\Z_2$. Let us define
\begin{equation}
\eqalign{
  x&=\xi_1^2,\cr y&=\xi_2^2,\cr z&=\xi_1\xi_2.\cr}
\end{equation}
Clearly $x,y,z$ are invariant under the $\Z_2$ action. We also
have $xy=z^2$. In fact this defines a one-to-one map between the space
$\C^2/\Z_2$ and the hypersurface $xy=z^2$ in $\C^3$. The latter forms
the description of the quotient singularity in algebraic geometry. Let
us pretend that $[x,y,z]$ are the homogeneous coordinates of $\P^2$
rather than the affine coordinates of $\C^3$. In this case $xy=z^2$
defines a smooth $\P^1\subset\P^2$. Putting coordinates $[a,b]$ on
this $\P^1$ we may map it into $\P^2$ via $x=a^2, y=b^2, z=ab$. The
definition of $\O(-1)$ amounted to taking the subspace of
$\P^1\times\C^2$ where the affine coordinates on $\C^2$ represented
the homogeneous coordinates on $\P^1$. Let us now play the same game
with our subspace of $\C^3$. That is we consider
\begin{equation}
  \{[a,b],(x,y,z)\in
\P^1\times\C^3;\,xy=z^2,a^2z=abx,a^2y=b^2x,aby=b^2z\}.
\end{equation}
Now, with reasoning similar to before, away from $(0,0,0)\in\C^3$ the
space looks like $\C^2/\Z_2$ whereas at $(0,0,0)\in\C^3$ we have a
$\P^1$. That is, we have replaced the singular point of $\C^2/\Z_2$ by
a $\P^1$ and in the process we have ended up with a smooth space!
Actually the above form of this space can be simplified. The
constraints are sufficient to uniquely determine $z$ from the other
variables. Thus the space is isomorphic to
\begin{equation}
  \O(-2)=\{[a,b],(x,y)\in\P^1\times\C^2;a^2y=b^2x\}.
\end{equation}
In general the notation $\O(n)$ represents the line bundle over
$\P^1$ whose first Chern class, when integrated over the base, equals
$n$. The condition $K=0$ is only met by $\O(n)$ when $n=-2$. Thus, it
is only when blowing up with $\O(-2)$ that we can hope to obtain a
\CY\ space in two dimensions.

Note that blowing-up changes the topology. In particular, introducing
the new $\P^1$ adds an element
to $H_2(X)$.
This new element is called an ``exceptional divisor''. The term
``divisor'' means, for our purposes, any linear combination of complex
codimension one objects in our algebraic variety. ``Exceptional''
refers to the fact that it came from a blow-up.
We may take our space obtained by dividing the complex
2-torus by $\Z_2$ and obtain a smooth space by blowing up all 16 fixed
points. The Hodge numbers $h^{p,q}$ can be calculated by starting with
the cohomology of the complex 2-torus, projecting out those elements
which are not invariant under the $\Z_2$ action and then adding in the
elements from the exceptional divisors of the blow-up. This is
captured by the following Hodge diamonds:
\newcommand{\m}[1]{\multicolumn{2}{c}{#1}}
\newcommand{\mapwith}[1]{\smash{\mathop{\longrightarrow}\limits^{#1}}}
\newbox\szbox
\begin{equation}
  \setbox\szbox=\hbox{$h^{3,3}$}
  {\arraycolsep=0.3\wd\szbox
  \begin{array}{*{6}{c}}
    &&\m1&& \\ &\m2&\m2& \\ \m1&\m4&\m1 \\
    &\m2&\m2& \\ &&\m1&&
  \end{array} \mapwith{/\Z_2}
  \begin{array}{*{6}{c}}
    &&\m1&& \\ &\m0&\m0& \\ \m1&\m4&\m1 \\
    &\m0&\m0& \\ &&\m1&&
  \end{array} \mapwith{{\rm Blow-up}}
  \begin{array}{*{6}{c}}
    &&\m1&& \\ &\m0&\m0& \\ \m1&\m{20}&\m1 \\
    &\m0&\m0& \\ &&\m1&&
  \end{array}
}
\end{equation}
The diamond on the left is that of the complex 2-torus and on the
right that of the desired space. The smooth \CY\ surface is an example
of a K3 surface. Any smooth complex surface which is \CY\ and is not a
torus will be diffeomorphic to a K3 surface (see for example
\cite{BPV:}). One may divide complex 2-tori by many discrete groups.
So long as this discrete group is a subgroup of $SU(2)$ one may
blow-up the resulting space and, if there were any fixed points, one
will find the same Hodge diamond as above. Blow-ups of
complicated quotient singularities are performed by a sequence of
adding $\O(-2)$ spaces \cite{BPV:}.

Let us now consider the K\"ahler form on $X$, when $X$ is a K3
surface. Then since $h^{1,1}=20$, the cohomology class of the K\"ahler
form lives in the vector space $\R^{20}$.
There is a natural linear mapping between $(1,1)$-forms and divisors.
This mapping is ``the
dual of the dual'' in the following sense. We take a $(1,1)$-form $e_i$
and find the dual form $\tilde e_i$ such that $\int_X \tilde e_i
\wedge e_j=\delta_{ij}$. $C_i$ is then taken to be the dual of $\tilde
e_i$ in the sense that $\int_{C_i}\tilde e_j=\delta_{ij}$.
Let us choose a basis $e_i$,
$i=1,\ldots,20$. One choice could be to take $e_{17},\ldots,e_{20}$
forming a basis
of the $(1,1)$-forms of the original two-torus and have
$e_1,\ldots,e_{16}$ associated to the 16 exceptional divisors. We then write
\begin{equation}
  J=\sum_{i=1}^{20}=J_i e_i.
\end{equation}
Using standard K\"ahler geometry (see, for example, \cite{GH:alg}) one
may calculate the area of a curve, $C$, given its homology class,
i.e., Area $=\int_CJ=\int_X e\wedge J$. To calculate this we
need to know the intersection form $\langle e_i,e_j\rangle=\int_X
e_i\wedge e_j$.
Our association between $e_i$ and $C_i$ allows us to rephrase
questions in terms of intersection numbers since
$\langle e_i,e_j\rangle= \#(C_i\cap C_j)_X$.

Let us consider the
intersection numbers of an exceptional divisor, say, $C_1$, with the
other generators. Each exceptional divisor comes from a different fixed
point and so need not touch the others. Also a generator of $H_2$ on
the original complex two-torus may defined such that it does not pass
through the fixed points. Thus $\#(C_i\cap C_1)_X=0$ for $i=2,\ldots,20$.
What about $\#(C_1\cap C_1)_X$? Consider the rational curve, $C$, in $\O(n)$
defined as the base space while considering $\O(n)$ as a line bundle.
This bundle may be considered as the normal bundle of the embedding of
this rational curve in the space $\O(n)$. Deformations of the curve
may then be given by holomorphic sections of this normal bundle. If
$n\geq 0$ there are non-zero sections of the bundle with generically $n$
zeroes. In this manner, we see that $C$ has self-intersection $n$. In
our case $n=-2$ and there are no deformations of the rational curve.
However one may assume that the above reasoning extends for $n<0$ and
we declare that $\#(C_1\cap C_1)_X=-2$. We therefore obtain
\begin{equation}
  \eqalign{
  \hbox{Area}(C_1)&=\int_{C_1}J\cr
	&=-2J_1.\cr}
\end{equation}

Since it is a reasonable assertion that the area of this curve should
be positive, we should therefore impose that $-J_1>0$. When one
computes all the areas and volumes of all algebraic subspaces of $X$
as well as the volume of $X$ itself, this positivity condition marks
out a region of $\R^{20}$ where the K\"ahler is allowed to have
values. The shape marked out is a cone --- if $J$ is a valid K\"ahler
form then so is $\lambda J$ for $\lambda$ a positive real number. This
subspace of $\R^{20}$ is called the ``K\"ahler cone''.

Note that when we take the limit $J_1\to0$, the area of the
exceptional divisor shrinks down to zero. This is precisely the
reverse of blowing-up. By sending $J_i\to0$ for $i=1,\ldots,16$ we can
thus recover the orbifold of the complex 2-torus. In this sense, the
orbifold ``lives'' in the boundary of the K\"ahler cone. This is shown
in figure \ref{fig:kc}.

\iffigs
\begin{figure}[t]
  \centerline{\epsfxsize=11cm\epsfbox{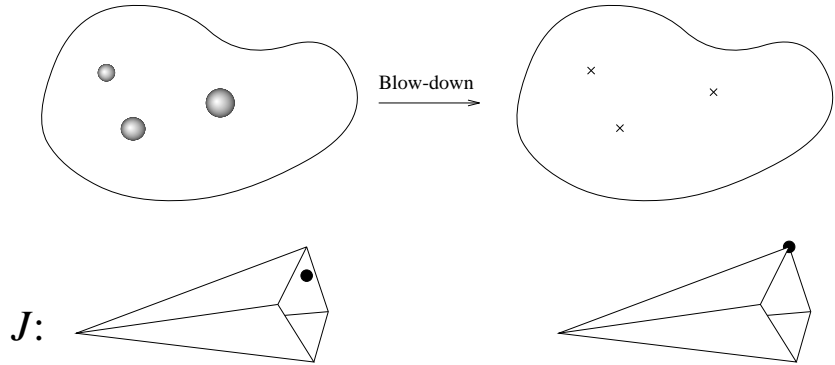}}
  \caption{Blowing down K3 and the associated K\"ahler form.}
  \label{fig:kc}
\end{figure}
\fi

It is worth mentioning that one may analyze the orbifold directly in
string theory without having to resolve the singularities
\cite{DHVW:}. To study the orbifold in $A/G$, where $A$ is a smooth
space and $G$ is a discrete group acting with fixed points, one considers
string theory on $A$. As well as closed strings on $A$ one should
also consider strings which are open but whose ends are identified
under the action of $G$ since such strings will be closed on the
orbifold. These are called ``twisted'' strings. One then projects on to all
$G$-invariant states. This procedure allows one to deduce the string
spectrum, and hence Euler characteristic, of the orbifold
\cite{DHVW:}. Applying such a method for our two-torus divided by
$\Z_2$ we obtain an Euler characteristic of 24 in agreement with the
cohomology of K3. This calculation is actually more surprising than
first meets the eye. The orbifold itself, despite being singular may
still be triangulated and thus singular homology may be defined on it.
The orbifold is a simplicial complex obtained from the two-torus by
removing 16
points, dividing by a freely acting $\Z_2$ and adding back the 16
points. The Euler characteristic is thus 8. Thus string theory somehow
knew that when it was on the orbifold that the associated geometry was
that of a (blown-down) K3 surface rather than the obvious simplicial
complex for purposes of cohomology. This is another reason to believe
that algebraic geometry is the best setting for understanding the
geometry of string theory --- blowing up the orbifold to form K3 is a
very natural process in algebraic geometry. While the correct geometrical
picture is now well-understood for simple cases (see, for example
\cite{me:orb2}) addition of so-called ``discrete torsion'' into the
orbifolding process can introduce many curiosities \cite{VW:tor}.

Now let us discuss the deformations of complex structure. Deformations
of a complex manifold are given by elements of $H^1(X,T)$, i.e.,
one-forms with values in the holomorphic tangent bundle.
If $X$ is a \CY\ manifold of $d$ dimensions, there is a unique
$(d,0)$-form (up to rescaling) which may be used to form an
isomorphism between $H^1(X,T)$ and $H^{d-1,1}(X)$. See \cite{GSW:book}
for a fuller explanation.
It is known
\cite{Tian:def} that all elements of this cohomology group give
deformations assuming $X$ is smooth. Thus the dimension of the moduli
space of complex structures is given by the Hodge number $h^{d-1,1}$
which in this case is 20. It is actually rather tricky to explicitly
show all 20 deformations in one model. Let us build K3 as an algebraic
variety as we did for the elliptic curve. We know that any \CY\
surface we can build that is simply connected will be a K3 surface.
One can show that the hypersurface in $\P^3$ defined by an equation of
total degree 4 in the homogeneous coordinates $[x_0,x_1,x_2,x_3]$ has
the required properties (see chapter 2 of \cite{Hub:book} for a
thorough account of such constructions). It is not too difficult to
show that writing down the most general quartic defining equation and
accounting for linear redefinitions of the coordinates leads to 19
deformations of the defining equation. Thus we appear to ``miss'' one of
the the deformations when we analyze the K3 surface in this way.

The reason for this is not too difficult to see. It comes from the
fact that the deformations of complex structure and those of the
K\"ahler form tend to ``interfere'' with each. Suppose we have an
algebraic curve $C$ embedded in our K3 surface. To such a curve we
associate a two-form $e(C)$ by the linear map discussed earlier. This
two-form will be a (1,1)-form by standard results in complex geometry
\cite{GH:alg}. Since $C$ is an element of $H_2(X,\Z)$, $e(C)$ will
also be an element of $H^2(X,\Z)$. As we vary the complex structure of
$X$, the way that the lattice $H^2(X,\Z)$ sits within the Hodge
decomposition $H^2(X,\C)\cong H^{2,0}\oplus H^{1,1}\oplus H^{0,2}$ will
vary. Thus, the very existence of such a curve $C$ will put
constraints on the complex structure. Our K3 surface does indeed have
such a curve --- if we define $\P^2\subset\P^3$ by the vanishing of a
generic equation linear in the homogeneous coordinates, then the
intersection of this ``hyperplane'' $\P^2$ with the quartic
hypersurface will be an
algebraic curve (of genus 3). The fact that there is only class of
such curves puts one constraint on the complex structure and lowers
the dimension of the moduli space from 20 to 19. In fact the general
rule is that if we have $n$ algebraic curves in the
variety generating $n$ dimensions of the K\"ahler cone, we only have
$20-n$ deformations of complex structure. Thus it is hopeless to try
to capture the whole moduli space this way.

The actual analysis of the moduli space of K3 surfaces using mirror
symmetry is very interesting but we will not pursue it much further
here since it is of little relevance to the rest of these lectures. The
reader is referred to \cite{AM:K3m,AM:K3p} for further details. What
we will present here is the final answer. It turns out the the moduli
space fails completely to factorize into complex structure times
K\"ahler parts. The form of the space turns out to be remarkably
similar to that of the torus:
\begin{equation}
  \cM_{\rm K3}\cong O(4,20;\Z)\backslash O(4,20)/(O(4)\times O(20)),
\end{equation}
where the indefinite metric being preserved is of the form
$(-E_8)\oplus(-E_8)\oplus H\oplus H\oplus H\oplus H$, where $E_8$ is the
definite even self dual lattice in 8 dimensions and $H$ is the
2-dimensional matrix $H_{ij}=0$ for $i=j$ and $H_{ij}=1$ for $i\neq
j$. From Narain's work \cite{N:torus} this is precisely the moduli
space for string on a left-right unsymmetric torus of dimension
(4,20). Why this should be the case is as yet only partially
understood \cite{Sei:K3}.

Note that the moduli space is again of the form of some smooth Teichm\"uller
space divided out by some modular group. Note also that this moduli
space includes orbifolds such as our complex two-torus divided by
$\Z_2$. It should be the case (but this has not yet been checked) that
the modular group should identify at least some of the orbifolds with
classically smooth models as predicted in \cite{Eg:K3}.


\section{Calabi-Yau Threefolds}		\label{s:d=3}

\subsection{Generalities}

Now let us progress on to the case $d=3$. Recall that for $d=1$ there
was only one \CY\ manifold --- the torus. For $d=2$ we add two
possibilities. How many \CY spaces are there for $d=3$? Despite
considerable effort by mathematicians in recent years, at this point
in time it is not known if the number is finite or even if there
are any bounds on the Euler characteristic! There are certainly many
methods of building \CY\ manifolds and all constructions which
magically produced a K3 surface (or occasionally a complex two-torus)
for $d=2$ tend to produce something different in each case when applied to
$d=3$. There are some constructions which can systematically produce a
very large number of \CY\ manifolds. Perhaps the most general
construction studied so far is that of complete intersections in toric
varieties \cite{Boris:m}. There are known \CY\ manifolds which cannot
be built this way however \cite{Sch:ell}. See \cite{Hub:book} for an
introduction to a few of these techniques.

There are some simple statements that can be made. Firstly all \CY\
spaces with $h^{2,0}\neq0$ are the complex 3-torus or a K3 surface
times a one-torus or some orbifold thereof \cite{Beau:CY}. We will
ignore such possibilities here although it should not be difficult
using the results of the preceding sections to analyze the moduli
spaces completely for these examples.
Making the assumption $h^{2,0}=0$ removes some of the complications
that occured in the previous section. The effect of the complex
structure on the Hodge decomposition $H^2(X,\C)\cong H^{2,0}\oplus
H^{1,1}\oplus H^{0,2}$ will now be trivial since $H^2(X,\C)\cong
H^{1,1}$. This allows us to factorize (at least locally away from the
boundary) the moduli space into deformations of complex structure and
deformations of the K\"ahler form.

Unfortunately we pay dearly
for this simplification. It was the $h^{2,0}=1$ property of the K3
surface that allowed us to build the $N$=4 superconformal algebra that
in turn controlled the perturbation theory for the non-linear \sm.
Surprisingly one receives no correction to the $\beta$ function for
the metric at two and
three loop with only $N$=2 supersymmetry but at four loops we obtain an
extra correction \cite{GVZ:4loop} for the metric:
\begin{equation}
  \beta_{i\bar\jmath}=-\frac1{2\pi}R_{i\bar\jmath}-\frac{4\zeta(3)}{3(4\pi)^4}
  \partial_i\partial_{\bar\jmath}\left[R_{k\bar lm\bar n}
  R^{p\bar lq\bar n}R_p{}^k{}_q{}^m-R_{k\bar lm\bar n}
  R^{m\bar np\bar q}R_{p\bar q}{}^{k\bar l}\right]+\ldots
	\label{eq:4lb}
\end{equation}
Thus we no longer have a Ricci-flat metric as a solution. Actually
this doesn't really matter since we never actually found the metric in
the last section, we just needed Yau's theorem. We now need some
modified version of the theorem to say that given $X$ with a complex
structure and a point in the K\"ahler cone of $X$, there will be a
unique solution of (\ref{eq:4lb}) (at least in some large radius
limit) such that the K\"ahler form is
of the desired cohomology class. We will assume that this is the case
for convenience of argument. Ultimately we do not really care about
the metric and this point will be irrelevant.

More worryingly, we now have instanton effects which are non-trivial.
As we shall see this will cause severe problems with a metric-based
point of view but fortunately we have the algebrao-geometric methods
of section \ref{s:torus} to rescue us.

The two factors of the moduli space are best viewed using topological
field theory. We refer to M. Bershadsky's lectures in this volume for
more details regarding the following summary. The main point is that a
topological field theory has only a few observables and correlation
functions which are fairly simple to calculate (at least compared to
non-topological field theories). We can ``twist'' an $N$=2 superconformal field
theory in two different ways to obtain two distinct topological field
theories. This is achieved via the transformation
\begin{equation}
  \eqalign{T(z)&\to T(z)\pm\ff12 \partial J(z),\cr
\bar T(\bar z)&\to \bar T(\bar z)+\ff12 \bar\partial \bar J(\bar
z).\cr}
\end{equation}
These two different twistings give rise to the ``A-model'' and the
``B-model'' as defined in \cite{W:AB}. Only the (anti)chiral
superfields of the $N$=2 theory appear in the A and B models. In
particular, only the (c,c) and (a,a) fields appear in one and the
(c,a) and (a,c) fields appear in the other (again see
\cite{BRG:lect} for an explanation of this notation).

When analyzed in terms of a non-linear \sm\, the correlation functions
of the fields in the A-model depend only upon the K\"ahler form and
$B$-term of $X$ and the correlation functions of the B-model depend
only upon the complex structure \cite{W:AB}. Thus our factors
of the moduli space of $N$=2 theories coincide with the moduli spaces
of the topological field theories obtained by twisting. Thus, although
considerable information is lost when one twists an $N$=2 theory into
a topological field theory, so long as we consider both twists, we do
not lose any information about the structure of the moduli space.

Let us briefly review each moduli space. The moduli space of the
A-model $\cM_A$ encodes information concerning the K\"ahler form and
$B$-term of $X$. This occurs because of instantons. A solution to the
equations of motion of the A-model is a holomorphic map from the
world-sheet, $\Sigma$, to $X$. The constant map mapping
$\Sigma$ to a point in $X$ is the trivial case. Any other solution is
an instanton. If this map is one-to-one then the image of $\Sigma$ in
$X$ is an algebraic curve. In the case $d=3$ it turns out that
three-point functions between interesting observables are only
non-zero when the world-sheet is genus zero. The curves in question
are therefore rational. The remaining instantons unaccounted for are
the non-trivial many-to-one maps. These are multiple covers of
rational curves (for which the interested reader is referred to
\cite{AM:rat}).

In the case that an instanton corresponds to a rational curve, $C$, in
$X$, the value of the action is $S=-2\pi i\int_C(B+iJ)$. Let us chose
a basis, $e_k$, for $H^2(X,\Z)$. Using our assumption that $h^{2,0}=0$
we may then make the following definitions
\begin{equation}
\eqalign{B+iJ&=\sum_k(B+iJ)_k\,e_k,\cr
  q_k&=e^{2\pi i(B+iJ)_k},\quad k=1,\ldots,h^{1,1}(X).\cr}
		\label{eq:qdef}
\end{equation}
It then follows that any correlation function will be a power series
in the variables $q_k$.

Note that we can now make more precise our rather vague notion of
``large radius limit'' for the non-linear \sm. As each component of
the K\"ahler form, $J_k$, tends to infinity, we see that $|q_k|\to0$.
Thus any power series is more likely to converge. We will define the
large radius limit to be the limit in which $J_k\to\infty$ for all
$k$. Thus, not only does $X$ become a space of infinite volume but
each algebraic subspace will also become infinitely large.
We will consider this large radius limit $q_k=0$ to be a point in the
moduli space. With this
definition we may hope that the correlation functions of the A-model
will converge in some non-zero region around this large radius limit point.

The B-model and its moduli space $\cM_B$ are in many ways much
simpler. The only solution to the
equations of motion in this case are the constant maps. Therefore we
are free from instantons. This makes the non-linear \sm\ very simple
to analyze for information concerning observables which appear in the
B-model. Essentially the three-point functions calculated at
tree-level in the large radius limit will be exact. Actually this fact
should have been clear once we noted the factorization of the moduli
space into $\cM_A\times\cM_B$. If the correlation functions of the
B-model depend only on our position with $\cM_B$, we can go as close
to the large radius limit in $\cM_A$ as we please thus taking the
classical limit. This has a profound consequence:

\[\fbox{\vbox{\em\noindent\hsize=13cm The moduli space of B-models, $\cM_B$,
associated to a \CY\
space $X$ is isomorphic to the classical moduli space of complex
structures on $X$.}}\]

Thus the fact that the left-hand side of figure \ref{fig:mod-t} is the
moduli space of complex structures for an elliptic curve was not an
artifact of the simplicity of this model. This kind of behaviour
persists in higher dimensions.

In the case of the complex one torus we just had to copy $\cM_B$ to
obtain $\cM_A$. In general this situation is not quite so
straight-forward. Let us consider a non-linear \sm\ with target space
$X$. We may then associate to this a conformal field theory and thus a moduli
space if $X$ is a \CY\ space. It may (or may not) be the case that
there is another \CY\ space $Y$ which yields exactly the same
conformal field theory as that given by $X$ except that we exchange
chiral rings (c,c)$\leftrightarrow$(a,c) and
(c,a)$\leftrightarrow$(a,a). In this case $Y$ is the ``mirror'' of $X$
(again the reader is referred to \cite{BRG:lect}.) If this is the
case then clearly $\cM_A(X)\cong\cM_B(Y)$ and $\cM_A(Y)\cong\cM_B(X)$.
For the cases $d<3$ the condition of being a \CY\ is so constraining
that $X$ and $Y$ are topologically the same. Thus the mirror map
appears as an automorphism on the moduli space. For $d\geq3$ it is
usually the case that $X$ and $Y$ are topologically distinct (or even
that $Y$ does not exists as \CY\ manifold).

Assuming $Y$ exists and we can identify it, we may thus calculate the
moduli space of $N$=2 theories associated to $X$ by two complex
structure moduli space calculations ---  one for $\cM_B(X)$ and one for
$\cM_A(X)\cong\cM_B(Y)$. The moduli space is then generically
$(\cM_A(X)\times \cM_B(X))/\Z_2$ where the final $\Z_2$ quotient
corresponds to complex conjugation of $X$ and changing the sign of
$B$.

\subsection{The Gauged Linear \sm}

Before we proceed further to look at a simple example, we first
introduce another field theory associated to the target space $X$.
The reason for this short diversion should become clear soon.
This will be Witten's linear \sm. We refer the reader to the original
paper \cite{W:phase} for more details concerning the remainder of this
subsection
(see also J. Distler's lectures in this volume). The basic idea is
that although the conformally invariant \sm\ for $X$ is very
complicated we may consider a much simpler \sm\ with a larger target
space containing $X$ such that the fields are constrained to live in $X$ in
some limit. The latter \sm\ (which we will refer to as the {\em
linear\/} \sm)
shares many properties of the nonlinear \sm\ which has $X$ as the
genuine target space.

Before writing down this linear \sm\ we need to specify our
conventions for superspace. To fit in with our earlier description of
the $N$=2 superalgebra we denote by $(z,\theta^\pm)$ our coordinates
for the left sector and $(\bar z,\bar\theta^\pm)$ for the right
sector.\footnote{Note that this differs from \cite{W:phase}.} Let us
introduce a set of charged (c,c) superfields $\Phi_i$ and a set of
neutral (c,a)
superfields $\Sigma_l$ in a supersymmetric gauge theory. That is,
\begin{equation}
  \cD_+\Phi_i=\barcD_+\Phi_i=\cD_+\Sigma_l=\barcD_-\Sigma_l=0,
\end{equation}
where $\cD$ is the gauge covariant superderivative. For our purposes
we consider the case where the gauge group is $U(1)^s$ and
$l=1,\ldots,s$. Each $\Sigma_l$ may be considered as the
supersymmetric version of the field strength for the \th{l}\ $U(1)$
factor \cite{W:phase}. The gauge fields for this theory live in
vector superfields $V_l$. We may write
\begin{equation}
  \Phi_i=\exp(\sum_{l=1}^s Q^{(l)}_iV_l)\Phi^0_i,
\end{equation}
where $Q^{(l)}_i$ is the charge of $\Phi_i$ with respect to the \th{l}\
$U(1)$. In this case $\Phi^0_i$ is a (c,c) superfield with respect to
the standard (rather than gauge covariant) superderivatives.
We may then expand $\Phi^0_i$ as
\begin{equation}
  \Phi^0_i=\phi_i+\psi\theta^-+\bar\psi\bar\theta^-+F_i\theta^-
    \bar\theta^-+\ldots
	\label{eq:spex}
\end{equation}
Note that for a rigid $U(1)$ rotation, $\Phi_i\to e^{i\alpha}\Phi_i$, the
vector superfield is invariant and so $\phi_i\to e^{i\alpha}\phi_i$.
The details of the vector superfield are rather messy. The
``$D$-term'' appears in the form
$D_l\,\theta^+\theta^-\bar\theta^+\bar\theta^-$ in the expansion. A vector
superfield in four dimensions has a real vector boson as its lowest
component. When we dimensionally reduce to two dimensions we thus
obtain a two-dimensional vector, which we denote $v_l$ and two real
scalars which we combine into a complex scalar $\sigma_l$.

The action to be considered is
\begin{equation}
  \eqalign{S=\int&\sum_{i=1}^N\left(\bar\Phi_i\Phi_i\right)\,d^2zd^4\theta
  -\frac1{4e^2}\int\sum_{l=1}^s\left(\bar\Sigma_l\Sigma_l\right)\,
	d^2zd^4\theta\cr
  &-\int W(\Phi_i)\,d^2zd^2\theta^-+\frac{i}{2\sqrt{2}}\sum_{l=1}^s
  (\beta+ir)_l\int\Sigma_l\,d^2zd\theta^+d\bar\theta^-
  +\hbox{h.c.},\cr}		\label{eq:linsig}
\end{equation}
where $W$ is a neutral holomorphic function and $r_l$ and $\beta_l$ are real
numbers.

The equations of motion for this model together with the condition
that we lie in the fixed point set of the fermionic
symmetries\footnote{This is where the topological field theory
localizes \cite{W:phase}.} give
\begin{equation}
\eqalign{D_l=-e^2\left(\sum_{i=1}^N Q^{(l)}_i|\phi_i|^2-r_l\right)&=-{\bf
F}_l\cr
  F_i=\frac{\partial W}{\partial \phi_i}&=0\cr
  \bar D\phi_i&=0\cr
  Q^{(l)}_i\phi_i\sigma_l&=0\cr}	\label{eq:eom}
\end{equation}
The solutions to these equations are governed by the
topological invariant
\begin{equation}
\int {\bf F}_l\,d^2z = -2\pi n_l,	\label{eq:c1F}
\end{equation}
where ${\bf F}_l$ is the field
strength of $v_l$. This is topological since it may be considered to
be the integral of the first Chern class over a space. Assuming single
valuedness of sections of the bundle of which ${\bf F}$ is the
curvature would force $n$ to be integer. This is the
two-dimensional analogue of the familiar second Chern class
$\int \mathop{\rm tr}(F\wedge F)\,d^4x$ term
that appears in four-dimensional Yang-Mills. One may also show
\cite{W:phase} that $\sum_lr_l n_l\geq0$.

The first and third equation of (\ref{eq:eom}) may be combined to
complexify the $U(1)^s$ gauge group to $(\C^*)^s$
\cite{W:phase,Brad:hol}.
($\C^*$ is the group, under multiplication, of nonzero complex numbers.)
We need only work at tree-level for the purposes of these lectures and
thus let us assume that $\Sigma$ is of genus zero.
The third
equation in (\ref{eq:eom}) may then be used to tell us that $\phi_i$ is a
holomorphic section of the line bundle $\O(\sum_l Q^{(l)}_in_l)$. The
only holomorphic section of $\O(m)$, where $m$ is negative, is the
trivial zero section and so we have
\begin{equation}
  \sum_{l=0}^sQ^{(l)}_in_l<0 \quad\Rightarrow\quad \phi_i=0.
	\label{eq:zero}
\end{equation}
This will force some, but in general not all, of the fields $\phi_i$ to vanish.
Therefore, generically, the last equation in (\ref{eq:eom}) forces
$\sigma_l$ to vanish.

For the classical vacuum we need to find the classical potential. This
is given by \cite{W:phase,MP:inst}
\begin{equation}
  U=\frac1{2e^2}\sum_{l=1}^sD_l^2+\sum_{i=1}^N|F_i|^2+2\sum_{k,l=1}^s
   \sum_{i=1}^N\bar\sigma_k\sigma_l Q^{(k)}_iQ^{(l)}_i|\phi_i|^2,
\end{equation}
where the equations of motion set the auxiliary fields as in (\ref{eq:eom}).

There is one very useful fact which we should immediately note
concerning the parameter $\beta_l$. This appears as the coefficient of
the term (\ref{eq:c1F}). $\beta_l$ behaves just like the ``theta'' in
four-dimensional Yang-Mills theory, which only affects the field
theory via instantons and correlations are periodic in this variable.
In particular, if $n$ is an integer then $\beta_l\cong \beta_l+1$.

\subsection{An example}		\label{ss:eg1}

Further discussion of the case $d=3$ and the linear \sm\
is best illuminated by a
specific example. From previous sections we have already discussed two
simple methods of
construction one might use to build a \CY\ threefold. The first method
from section \ref{s:K3t} would be to take a complex three-torus and
divide out by a subgroup of $SU(3)$ and blow-up the resulting
singularities. The other method
would be to take a complex projective space $\P^4$ with homogeneous
coordinates $[x_0,\ldots,x_4]$ and define a hypersurface, $X$, by the
vanishing of a generic polynomial of homogeneous degree 5 in $x_i$.
Both of these examples appeared in the original paper on \CY\
manifolds in string theory \cite{CHSW:}.

The latter space above, the ``quintic threefold'', is particularly
attractive because of the simplicity of $\cM_A$.
The moduli space $\cM_A(X)$ of this \CY\ space was computed first in
\cite{CDGP:}.
Taking a hyperplane
$\P^3$ and intersecting with the quintic threefold we obtain a
divisor. (Since $d=3$, divisors will now have complex dimension 2.)
This divisor generates
$H_4(X,\Z)$ and thus there is a two-form, $e$, associated (by the same
``dual of a dual'' construction of section \ref{s:K3t}) to this
divisor that generates $H^2(X,\Z)$. Thus the dimension of $H^2(X)$ is
one. The full Hodge diamond of this space is
\def\m#1{#1}
\begin{equation}
h^{p,q}(X) =
  {\arraycolsep=2pt
  \begin{array}{*{7}{c}}
    &&&\m1&&& \\ &&\m0&&\m0&& \\ &\m0&&\m1&&\m0& \\
    \m1&&\m{101}&&\m{101}&&\m1 \\
    &\m0&&\m1&&\m0& \\ &&\m0&&\m0&& \\ &&&\m1&&&
  \end{array}}.
\end{equation}

Therefore $\cM_A$ is has complex dimension 1 and $\cM_B$ has
dimension 101. Before discussing $Y$
and $\cM_B(Y)$ let us explore the linear \sm\ of $X$.
For our example we set $s=1$ and $N=6$. Write $\Phi_1,\ldots,\Phi_5$
as $X_0,\ldots,X_4$ and $\Phi_6$ as $P$. Also set
\begin{equation}
\eqalign{W&=P.f(X_i)\cr
  &=P(X_0^5+X_1^5+X_2^5+X_3^5+X_4^5+\ldots),\cr}
\end{equation}
where $\ldots$ represents other terms (with arbitrary coefficients) of
degree 5 in the $X_i$'s.
For $W$ to be neutral we set the $U(1)$ charge of each $X_i$ to be +1
and the charge of $P$ to be $-5$. We will also use the letters
$x_i$ and $p$ to denote the complex bosons which are the lowest
respective components of the associated (c,c) superfields in the sense of
(\ref{eq:spex}).
The classical potential energy becomes
\begin{equation}
  U=\left[|x_0|^2+\ldots+|x_4|^2-5|p|^2-r\right]^2 + |f(x_i)|^2+
		|p|^2\sum_{i=0}^4\left|\frac{\partial f}{\partial x_i}
		\right|^2+2|\sigma|^2\left(\sum_{i=0}^4|x_i|^2+
		25|p|^2\right).	\label{eq:thisU}
\end{equation}
We also have from (\ref{eq:eom}) that $x_i$ are sections of $\O(n)$
and that $p$ is a section of $\O(-5n)$. In addition, the fields
$\phi_i$ satisfy the constraint $f(x_i)=0$.

Suppose for the time being that $r\gg0$. Since $nr\geq0$, we have $n\geq0$.
If $n>0$ then $p$ must be zero by (\ref{eq:eom}).
Clearly $U$ is non-negative. Let us try to solve for $U=0$ --- this
will be the $n=0$ case. To obtain
zero for the first term in
(\ref{eq:thisU}) we must have that at least one of the $x_i$'s is
non-vanishing. The last term then forces $\sigma$ to vanish. Assuming
the transversality condition holds, i.e., that (\ref{eq:smooth}) is only
satisfied when all the $x_i$'s are zero, then $p$ is forced to vanish
by (\ref{eq:eom}).

The classical vacuum, i.e. $U=0$, is thus parametrized by $x_i$
subject to
the constraints
\begin{equation}
  \eqalign{|x_0|^2+\ldots+|x_4|^2&=r\cr
	f(x_i)&=0.\cr}  \label{eq:con1}
\end{equation}
We also have the $U(1)$-action $x_i\to e^{i\alpha}x_i$. The first
equation in (\ref{eq:con1}) gives the sphere $S^9$. Dividing this by
the $U(1)$ action gives $S^9/S^1\cong\P^4$. The easiest way to see
this latter isomorphism is to build $\P^4$ in stages as follows:
\begin{enumerate}
  \item Take the space $\C^5$ with coordinates $(x_0,\ldots,x_4)$ and
remove the origin $\{O\}$.
  \item Divide any vector in $\C^5-\{O\}$ by its length. This retracts
$\C^5-\{O\}$ onto the sphere $S^9$. It may be considered as the quotient
$(\C^5-\{O\})/\R_+\cong S^9$ where $\R_+$ is the group of positive real
numbers.
  \item Since $\P^4\cong (\C^5-\{O\})/\C^*$ and $\C^*\cong
S^1\times\R^+$, we have $\P^4\cong S^9/S^1$ where the $S^1$ acts as
$x_i\to e^{i\alpha}x_i$.
\end{enumerate}
 Finally, therefore, our target space consists of the hypersurface
$f(x_i)=0$ in the projective space $\P^4$ exactly as desired. It
should be noted that this manipulation extending a $U(1)$ quotient to
a $\C^*$ quotient is essentially equivalent to the procedure referred
to earlier for identifying $\phi_i$ as sections of holomorphic bundles.

Deforming the quintic equation $f(x_i)$ will produce deformations of
complex structure of $X$. More interestingly, changing $r$ will change
the radius of $S^9$ and thus the overall scale of $X$. This is
equivalent to deforming the single parameter governing the K\"ahler
form of $X$. Thus, $r$ appears to be a degree of freedom similar to
$J$. It is also tempting to associate $\beta$ with $B$ since they both
live on a circle and combine with $r$ and $J$
respectively to form natural complex parameters. Before we make such a
bold identification however we must bare in mind that we have not
really built the desired non-linear \sm\ with target space $X$.

The fields $x_i$ originally span the space $\C^5$. The metric on this
space is the trivial one (implicitly present in the first term in
(\ref{eq:linsig})). The metric on $\P^4$ is inherited from this
original metric by ``symplectic reduction'' \cite{Audin:}. This
produces the ``Fubini-Study'' metric on $\P^4$. The metric on $X$ is
therefore the restriction of the Fubini-Study metric. Such a metric is
not Ricci flat and so cannot satisfy (\ref{eq:4lb}) in the large
radius limit. Actually we have not even the correct degrees of freedom
in the linear \sm. In the desired nonlinear \sm\ the fields are
completely constrained to lie in the target space $X$. While we appear
to have the correct behaviour for the low-energy behaviour in the
linear \sm\ one should expect extra massive fields which are not
confined to live in $X$. Therefore, in order to obtain the correct
degrees of freedom we should integrate out such massive states. Such a
procedure may well affect the value of the parameters $\beta$ and $r$.

The correlation functions of the A-model depended on $B+iJ$. This
dependence came from instanton effects. To compare the quantities
$B+iJ$ and $\beta+ir$ it should be useful to look at instantons in the
linear \sm. The interested reader is referred to \cite{MP:inst} for
more features of these instantons. The instantons are solutions to the
equations (\ref{eq:eom}) but will not necessarily satisfy
$U=0$. In particular we have that $x_i$ are sections of $\O(n)$.
This forces $n$ to be an integer and thus $\beta\cong\beta+1$ as
explained in the previous section.

Now let us compare this linear \sm\ instanton with the A-model
instantons.\footnote{The author wishes to thank R. Plesser for
discussions on this point which has overlap with \cite{MP:inst}.} An
A-model instanton is a holomorphic map
from $\Sigma$
into $X$.
Homogeneity requires $n$ to be the same for each $x_i$. Now consider a
hyperplane $\P^3\subset\P^4$. The image of $\Sigma$ under the map
given by $x_i$ will intersect this hyperplane $n$ times if the map is
suitably generic. This is the {\em degree\/} of the map.

Thus far the linear \sm\ instantons and the A-model instantons appear
identical. There is a difference however. For the A-model, the
quantities $x_i$ are homogeneous coordinates of $\P^4$ and cannot
simultaneously vanish. The $x_i$'s of the linear \sm\ are just
sections of $\O(n)$ and therefore not so constrained.
Suppose we have an instanton in the linear \sm\ where all of the
fields $x_i$ vanish at a point $z_0\in\Sigma$ and let us assume that
$n=1$. The equations (\ref{eq:eom}) and (\ref{eq:c1F}) dictate that
\begin{equation}
 \int_\Sigma\left(\sum_{i=0}^4|x_i|^2-r\right)\,d^2z=\frac{2\pi}{e^2}.
\end{equation}
Since $r\gg0$, the value of $\sum_i|x_i|^2$ must rise rapidly as one
moves away from $z_0$. In fact, the region on the world sheet around
$z_0$ where $\sum_i|x_i|^2$ is not roughly equal to $r$ must have an
area the order of $1/e^2r$. That is, this instanton appears as a small
lump around $z_0$.

We see then that the linear \sm\ contains all the instantons of
A-model and in addition some small-scale instantons which shrink down
to points on the world-sheet in the limit $r\to\infty$. In order to
translate between the coordinate $r$ and the coordinate $J$ on the
respective moduli spaces we need to take into account the effects of
these small-scale instantons.

In addition to this non-perturbative correction to $r$ we should also
consider loops in the perturbation theory. The super-renormalizability
of this gauge theory in two dimensions makes such an analysis rather
straight-forward. We need to consider one-loop diagrams such as the
tadpole given by
\begin{equation}
\setlength{\unitlength}{0.012500in}%
\begin{picture}(150,60)(350,580)
\thinlines
\put(340,620){\line(1, 0){60}}
\put(420,620){\circle{40}}
\put(350,600){\makebox(0,0)[lb]{\smash{$D_l$}}}
\put(450,615){\makebox(0,0)[lb]{\smash{$\phi_i$}}}
\end{picture}	\label{eq:Dtad}
\end{equation}
This results in a shift in $\beta_l+ir_l$ which we will denote $\Delta_l$. The
value of this shift will vary depending upon our position in the
moduli space but should be expected to be finite.

Let us write $t_l=(B+iJ)_l$ and $\tau_l=(\beta+ir)_l$, where $r$ is
the {\em uncorrected\/} parameter in the linear \sm\ (i.e., we have
not taken into account the one-loop effects). To calculate
how $t_l$ is related to $\tau_l$ we combine the one-loop and instanton
corrections. This amounts to
\begin{equation}
  t_l = \tau_l+\Delta_l
	+\sum_{m=1}^s K_me^{2\pi i\tau_m}+\ldots \label{eq:shift}
\end{equation}
where $K_m$ represents the first-order effect from the small-scale
instantons and $\ldots$ represents the higher orders.
Let us introduce a variable $z_l$ as an analogy to the variable $q_l$
of (\ref{eq:qdef}):
\begin{equation}
  z_l = \pm\exp\left\{2\pi i\tau_l\right\}.
\end{equation}
The above sign ambiguity is a problem that always occurs in the
problem of trying to find the relationship between $q$ and $z$. The
idea is that it appears that one can choose a sign so that \cite{AGM:mdmm}
\begin{equation}
  q_l=z_l(1+C_l),		\label{eq:q=z}
\end{equation}
where $C_l$ is a power series in $z_1,z_2,\ldots$ with no constant term.
That is, we assume that the $\Delta$ term corrects $\tau_l$ by 0 or
$\pi$. One can determine this explicitly by counting rational curves
in the example being studied as was done in \cite{CDGP:} for
example. This conjecture has yet to be proven in general.
Standard
renormalization arguments in $N$=2 theories guarantee that $C_l$ is a
holomorphic function.

Thus far the reader may wonder what we have achieved by rephrasing
things in terms of the linear \sm. The answer is that we may now probe
the moduli space a long way from the large radius limit at least in
some sense --- the linear \sm\ may be analyzed away from the limit
$r\to\infty$, i.e., $|z|\to0$.

As an extreme example let us consider the case $r\ll0$.
Now $n\leq0$ and
the fields $x_i$ are forced to be zero (using the transversality
condition when $n=0$).
For the classical vacuum we first look at
the vanishing of $U$ in (\ref{eq:thisU}). This will force $p$ to be
nonvanishing. This in turn forces $\sigma$ to
vanish. The equation $f(x_i)=0$ is now trivially satisfied. The value
of $r$ fixes $|p|$ and the $U(1)$ gauge symmetry may be used to fix
the phase of $p$. Thus all the expectation values of the fields are
fixed --- the classical target space is a point! Actually, to be more
precise, we have a Landau-Ginzburg orbifold theory. The fields $x_i$
may have a zero vacuum expectation value but their quantum
fluctuations are
massless and governed by the superpotential $W$ which is of degree 5
in the fields $x_i$. We also need to note that the field $p$ has
charge $-5$ while the charge of the $x_i$'s is 1. Therefore, when the
phase of $p$ was fixed by using the symmetry $\phi\to
\exp(2\pi i\alpha Q_i)\phi$, we still have a residual $\Z_5$ symmetry
\begin{equation}
  g\colon x_i\mapsto e^{2\pi i/5}x_i.	\label{eq:Z5}
\end{equation}
Therefore, the fields $x_i$ effectively live in $\C^5/\Z_5$ and our
theory is an orbifold.

Again this theory has instantons. Now $p$ will be nontrivial and a section of
$\O(-5n)$. This means that $n\in\Z/5$ but that $n$ need not be
integral. Let us consider the instanton given by $n=-\ff15$. $p$ is a
section of $\O(1)$ and will have one zero on $\Sigma$. Let us denote this
point on the world-sheet by $z_0$.
Since
\begin{equation}
  \int_\Sigma e^2(-5|p|^2-r)\,d^2z=\int_\Sigma{\bf F}d^2z=\ff15,
\end{equation}
we must have that the value of $|p|^2$ rises quickly to $|r/5|$
outside a patch of area of order $-1/e^2r$ around this zero on the
world-sheet. Thus the interesting part of this instanton configuration
is confined to a small lump around this single zero. The fields $x_i$
live in the bundle $\O(-1/5)$ and will not be single valued. In
particular, the field $p$ will pick up a phase $2\pi$ with respect to
monodromy around $z_0$ whereas $x_i$ will pick up a phase of
$-2\pi/5$ around the same point $z_0\in\Sigma$. This is precisely a
twist-field configuration where the map $x_i$ is taking the point
$z_0$ on the world-sheet to the origin of the target $\C^5/\Z_5$.

Thus by varying the value of $r$, we may switch
between a target space of a \CY\ hypersurface in $\P^4$ (for which the
only massless modes lie within that space) and a target space which is a
point with massless Landau-Ginzburg-type massless fluctuations about
it. Each of these theories has instantons and in each case the action
of the instantons goes as $|r|$ so that the instanton effects become
negligible in the large $|r|$ limit. That is, we have ``exactly'' a
\CY\ theory for $r=\infty$ (in an infinitely large target space) and
``exactly'' a Landau-Ginzburg orbifold theory for $r=-\infty$.

We should ask if anything peculiar happens for a finite value of $r$
as we change between these two ``phases''. The only special value at
which something nasty happens occurs classically for $r=0$. At this
value, all the fields $x_i$ and $p$ may vanish and then $\sigma$ may
take on any value. Any analysis which works for the theory at generic
$r$ values should be expected to contain divergences when $r$ takes on
this value and $\sigma$ becomes massless.
Actually we have to be a little more careful than this (and we refer
the reader again to \cite{W:phase} for more detail).  Taking quantum
effects properly into account, the minimum energy density for a state
at large $\sigma$ goes roughly as $r^2 + \beta^2$. Thus
we have a singularity at $\beta+ir=0$.
This is the true corrected value for $\beta$ and $r$ at the
singularity so we need to compute $\Delta$ at this point in the moduli
space to find $z$. This
may be calculated by assuming that $\sigma$ is large and is given
purely by the diagram (\ref{eq:Dtad}). The result in
the case that $s=1$ is that \cite{MP:inst}
\begin{equation}
  \eqalign{\Delta&=-\frac1{2\pi}\sum_{i=1}^N Q_i\log Q_i\cr
	&=\frac5{2\pi}\log5-\ff52i.\cr}
\end{equation}
Thus $z=\pm5^{-5}$ at the singularity. It turns out that we should
choose $z=5^{-5}$ to get the rational curve count correct
\cite{CDGP:}.

To summarize thus far we have identified three special points in the
moduli space. At $z=0$ we have a \CY\ target space of infinite
radius. We also know that
in the neighbourhood of $z=0$ we may use (\ref{eq:q=z}) to relate this
coordinate to the coordinate in the moduli space of A-models on the
same \CY\ target space. At $z=\infty$ the linear \sm\ is equivalent to
a Landau-Ginzburg orbifold and at $z=5^{-5}$ the theory is singular.

Now let us address the question of $\cM_B(Y)$ which we expect to be
isomorphic to $\cM_A(X)$. First let us carefully build $Y$. Analysis
of the linear \sm\ showed us that there is a Landau-Ginzburg orbifold
theory in the moduli space. The same must be true for the theory of
A-models (although at this point in the argument we don't know where
it is located) and therefore $N$=2 conformal field theories. If we go
to the correct point in $\cM_B(X)$, the
Landau-Ginzburg theory in question may be written as a tensor product
of minimal $N$=2 models and thus the required Landau-Ginzburg orbifold
is an orbifold of such a tensor product \cite{VW:,Mart:,GVW:}. The
resulting theory is a ``Gepner Model'' \cite{Gep:}. The process for
finding the mirror of a Gepner model is well-understood \cite{GP:orb}
(see also \cite{BRG:lect}). The result is that the
mirror is a certain orbifold of the original Gepner model. This may
then be retranslated back into Landau-Ginzburg orbifold language.

The required result is as follows. If $\Gep{X}$ is the Landau-Ginzburg
theory defined by the superpotential\footnote{From this point onwards
we will use lower case to refer to chiral superfields as well as their
lowest bosonic component. We hope the context makes it clear which is
relevant.}
\begin{equation}
  W=x_0^5+x_1^5+x_2^5+x_3^5+x_4^5
\end{equation}
in $\C^5/\Z_5$ where $\Z_5$ is generated by (\ref{eq:Z5}), then the
mirror theory $\Gep{Y}$ is the Landau-Ginzburg orbifold theory with
the same superpotential in the space $\C^5/(\Z_5)^4$, where $(\Z_5)^4$
is the group consisting of elements
\begin{equation}
  g\colon(x_0,x_1,\ldots,x_4)\mapsto(\alpha^{n_0}x_0,\alpha^{n_1}x_1,
		\ldots,\alpha^{n_4}x_4),
\end{equation}
where $\alpha$ is a non-trivial fifth root of unity and $n_i$ are
integers such that the relation $\sum_i n_i=0\pmod5$ holds.

Now we need to look at $\cM_B(Y)$. This is given by the moduli space
of superpotentials. In the case of $\cM_B(X)$ there was a 101
dimensional space of superpotentials, or, equivalently, a 101
dimensional space of complex structures on $X$. The $(\Z_5)^4$
orbifolding group will project most of these deformations out. As
expected, we are left with one parameter. We may write $f$ in the
form
\begin{equation}
  f=x_0^5+x_1^5+x_2^5+x_3^5+x_4^5-5\psi x_0x_1x_2x_3x_4,
		\label{eq:Ydef}
\end{equation}
where $\psi$ is the parameter. Thus, by construction, $\psi=0$ is the
Landau-Ginzburg orbifold point which is mirror to the Landau-Ginzburg
orbifold point for $X$. Varying $\psi$ will span the space $\cM_B(Y)$.
At this point $Y$ is a Landau-Ginzburg orbifold. As we found with $X$
however, the space of $\cM_A(Y)$ will be such that $Y$ may be pictured
as a smooth \CY\ manifold in part of the moduli space of A-models.
Since we are only concerned with B-model data for $Y$ we may take $Y$
to be this smooth \CY\ manifold. As a matter of fact there will be many
smooth \CY\ manifolds which may be used to represent $Y$. This issue
will be explained further in section \ref{s:phase}.

Since we now have a global description of $\cM_B(Y)$ and of linear \sm
s on $X$, it is natural to try to map them to each other. First we
should pick good coordinates on each moduli space such that two
different values of the coordinate do not correspond to the same
point.
The identification $\beta\to\beta+1$ in the neighbourhood of $z=0$
makes $z$ a good coordinate at least in the vicinity of the large
radius \CY\ limit. Around the Landau-Ginzburg point, the instanton
number $n$ need only be an element of $\Z/5$ and so it would appear
that the identification should be $\beta\to\beta+5$. Actually there is
a $\Z_5$ symmetry around the point $z=\infty$ which identifies theories
such that we restore $\beta\to\beta+1$. As indicated earlier, the
instantons with non-integer $n$ may be thought of as twist fields. A
twist field $\xi_n$ then transforms under this $\Z_5$ symmetry as
\begin{equation}
  g\colon\xi_n\to e^{2\pi in}\xi_n.
\end{equation}
Therefore $z$ is a good coordinate locally around $z=\infty$.
Picturing the moduli space as $\P^1$ with $z$ the usual affine
coordinate, we have done enough to show that $z$ is a good coordinate
everywhere (together with the point $z=\infty$).

The $\Z_5$ transformation generated by
\begin{equation}
  g\colon(x_0,x_1,\ldots,x_4)\mapsto(\alpha x_0, x_1,\ldots,x_4)
\end{equation}
can be made a symmetry of (\ref{eq:Ydef}) by extending it so
that $g\colon\psi\to\alpha^{-1}\psi$. Therefore $\psi$ is not a good
coordinate. This symmetry of $\psi$ is the only one induced by
symmetries of the equation (\ref{eq:Ydef}). This means that $\psi^5$ is a
good coordinate.

We now wish to map these two $\P^1$ moduli spaces to each other in a
one to one manner. The pervading $N$=2 structure present means that we
expect this map to be holomorphic. This forces
\begin{equation}
  z=\frac{a\psi^5+b}{c\psi^5+d}
\end{equation}
where $a,b,c,d$ are complex numbers such that $ad-bc=1$. We may thus
fix this map by identifying three points. We have already identified
$z=\infty$ and $\psi=0$ as the Landau-Ginzburg orbifold point. We know
that $z=5^{-5}$ corresponds to a singular theory. The equation
(\ref{eq:Ydef}) satisfies (\ref{eq:smooth}) when $\psi^5=1$.
This therefore provides a natural candidate for the singular point in
$\cM_B(Y)$. Actually, if one attempts to calculate correlation
functions using the ``chiral ring'' of \cite{VW:} then it is precisely
when (\ref{eq:smooth}) is satisfied that these calculations become
badly defined. Thus we map $z=5^{-5}$ to $\psi^5=1$. For our last point we
pick the large radius \CY\ point $z=0$. Clearly it would be unnatural
for this point to be anything other than $\psi=\infty$. To check this
however one may put the Zamolodchikov metric on $\cM_A(X)$ and
$\cM_B(Y)$. The point $z=0$ maps to $q=0$ in $\cM_A(X)$ which
corresponds to the large radius limit and is thus an infinite distance
away from other points on $\P^1$. The same is true for $\psi=\infty$
on $\cM_B(Y)$ \cite{CDGP:}. We have thus fixed
\begin{equation}
  z=(5\psi)^{-5}.  		\label{eq:zpsi}
\end{equation}

It is amazing just how simple (\ref{eq:zpsi}) is. This is the justification
of our using the linear \sm. While the map between $\cM_A(X)$ (in
terms of $q$) and
$\cM_B(Y)$ (in terms of $\psi$) is rather complicated, there is a very
simple relation between the linear \sm\ version of $\cM_A(X)$
(in terms of $z$) and $\cM_B(Y)$. At this point it is very tempting to
end our analysis of the moduli spaces since in many ways we have all
the information we need concerning the moduli space. The only problem
is that we have described $\cM_A(X)$ in terms of the coordinate $z$
rather than $q$. Recall that $q$ was derived from the K\"ahler form
and thus ultimately differential geometry. It appears that $z$ is the
coordinate that appears more naturally in string theory and perhaps we
should rewrite general relativity in terms of degrees of freedom
naturally expressed in terms of this parameter rather than $q$.
Rather than attempting such an ambitious problem we will submit to the
present conventions for describing space-time and try to reparametrize
our moduli space in terms of $q$.

Actually the analysis of section \ref{ss:t-alg} allows us to do this
without much more work. Recall that we found $(B+iJ)_l$ in terms of a
ratio of periods on the mirror, in this case $Y$. One may use the
local geometry of the moduli space $N$=2 theories to show that this
is also the case for any $d$ \cite{BCOV:big}. Therefore we just need
the Picard-Fuchs equation for these periods to find $q$ as a function
of $z$. The desired equation for the periods $\varpi$ is \cite{CDGP:}
\begin{equation}
  \left(z\frac d{dz}\right)^4f-z\left(z\frac d{dz}+\ff15\right)
	\ldots\left(z\frac d{dz}+\ff45\right)f=0.
\end{equation}
The four solutions of this hypergeometric differential equation are
characterized by their monodromy around $z=0$, which goes as
$1,\log z,(\log z)^2,(\log z)^3$ respectively. The powerful constraint
(\ref{eq:q=z}) then determines the answer uniquely to be
\begin{equation}
  q=\exp\left(\frac{\varpi_1}{\varpi_0}\right),
\end{equation}
where
\begin{equation}
  \eqalign{\varpi_0&=\F43(\ff15,\ff25,\ff35,\ff45;\,1,1,1;\,5^5z)\cr
	&=\sum_{n=0}^\infty\frac{(5n)!}{(n!)^5}z^n\cr
  \varpi_1&=\varpi_0.\log z+5\sum_{n=1}^\infty
	\frac{(5n)!}{(n!)^5}z^n
	\left[\Psi(1+5n)-\Psi(1+n)\right],\cr}
\end{equation}
where $\Psi(x)$ is the digamma function defined as the derivative of
$\log\Gamma(x)$. These periods may also be found directly from a
smooth model of $Y$ without going via the Picard-Fuchs equation
\cite{lots:per}.

We may now expand $q$ as
\begin{equation}
  q = z + 770z^2+1014275z^3+\ldots	\label{eq:q=zquin}
\end{equation}
which converges for $|z|\leq5^{-5}$. Thus, by using mirror symmetry we have
been able to determine the precise form of (\ref{eq:q=z}). It should
come as no surprise that this series has a radius of convergence equal
to $5^{-5}$ since the singular theory lies at $z=5^{-5}$. At the point
$z=5^{-5}$ the
series just manages to converge and we obtain
\begin{equation}
  (B+iJ)_1 \approx 1.2056i.
\end{equation}
We denote this value of $J_1$ by $J_0$.
If we have a \CY\ manifold diffeomorphic to the quintic
hypersurface and if the size of it is such that the area of a curve
which generates $H_2(X)$ is larger than about
$1.2(4\pi^2\alpha^\prime)$ then we can place precisely where our model
lies in the moduli space of $N$=2 theories by using
(\ref{eq:q=zquin}).

We have not yet accounted for the complete moduli space of these
theories however. What about the region where $|z|>5^{-5}$? At this
point the series fails to converge. If we try to compute
correlation functions using the nonlinear \sm\ for a point in this
region of the moduli space we will also find that the instanton
corrections form a divergent series too. If we were to be pragmatic at
this point we should say that the nonlinear \sm\ is not such a good
picture for any theory in the region $|z|>5^{-5}$ and that some other
picture is more appropriate. Indeed, we already know what the other
picture is --- we should interpret such theories as Landau-Ginzburg
orbifolds corrected by instantons. Actually, to use more conventional
conformal field theory language, these instantons are twist fields.
One may deform a conformal field theory orbifold by a twist field
which is also a truly marginal operator. Although the resulting theory
will not generically be an orbifold, one may compute correlation
functions as a power series in terms of the fields in the original
orbifold theory. This will be a power series in the coupling to the
marginal operator in the usual way in conformal perturbation theory.
One will find that such a series will converge when $|z|>5^{-5}$.

For any point (except when $|z|=5^{-5}$) in the moduli space
$\cM_A(X)$ we may therefore associate an effective target space
theory. In one region we have a \CY\ manifold with instanton corrections
and in the other region we a Landau-Ginzburg orbifold perturbed by
twisted marginal operators. Both types of theories have correlators in
the form of convergent power series. The regions do not overlap.

This leads to the ``phase'' picture of the moduli space
\cite{W:phase,AGM:II}. We do not have some universal picture for a
target space $X$ associated to $\cM_A(X)$. It is necessary to have a
set of target space descriptions which are applicable in various
regions in $\cM_A(X)$.

The attentive reader may have realized that, because we have
holomorphic functions, we should be able to analytically continue our
series beyond their radius of convergence. Certainly this is possible
as we now explain. One can therefore begin to interpret one phase of
the moduli space in terms of other phases. The more cautious reader
might declare such a process to be unnatural however. Anyway, let us
see what happens when we extend the \CY\ picture into the
Landau-Ginzburg region.

To analytically continue we first need to apply branch cuts.
The branch points are clearly $z=0,5^{-5},\infty$. In our
identification of the strip $(B+iJ)_1$, where $0\leq B_1<1, J_1\geq
J_0$, with the
hemisphere $|z|\leq5^{-5}$ we have already cut from $z=0$ to
$z=5^{-5}$. We require another cut emanating from $z=\infty$. To avoid
changing the hemisphere around $z=0$ which is already in the desired
form, we cut form $z=\infty$ to $z=5^{-5}$. This latter cut may
be performed by returning to our variable $\psi$ and imposing
\begin{equation}
  -\frac{2\pi}5<\arg\psi<0.
\end{equation}
The same method of Barnes integrals as was used in section
\ref{ss:t-alg} then yields
\begin{equation}
  (B+iJ)_1=\ff12 + \ff{i}2\left\{\cot\ff\pi5+
  \frac{\Gamma^4(\ff45)\Gamma(\ff25)}{\Gamma(\ff15)\Gamma^4(\ff35)}
  (\cot\ff\pi5-\cot\ff{2\pi}5)e^{\frac{\pi i}5}\psi
  +O(\psi^2)\right\},
\end{equation}
as a convergent power series for $|\psi|<1$.
We may now map the entire $\P^1$ of $\cM_A(X)$ into the
$(B+iJ)_1$-plane. The result is shown in figure \ref{fig:BJquin}.

\iffigs
\begin{figure}[t]
  \centerline{\epsfxsize=13cm\epsfbox{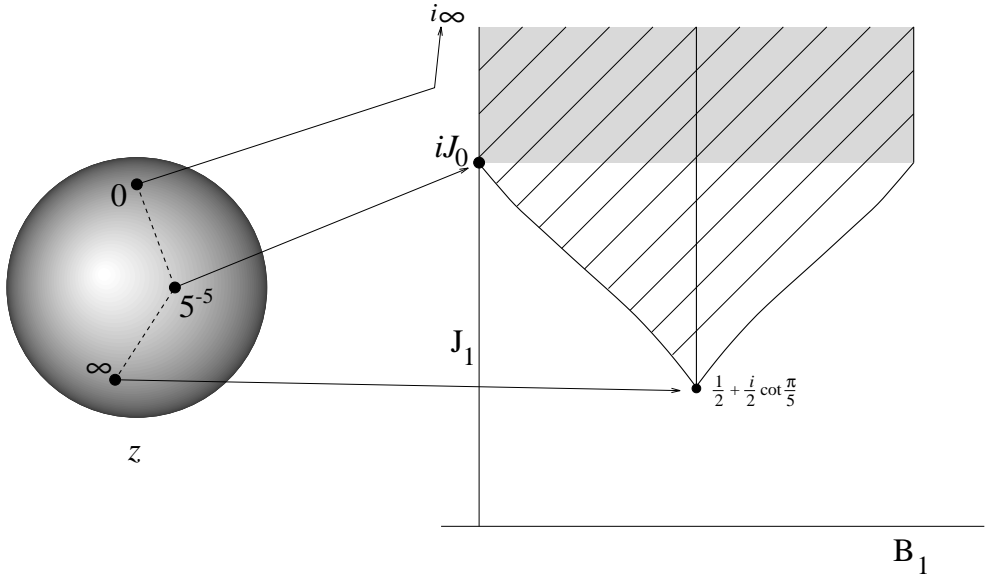}}
  \caption{Mapping $\cM_A(X)$ to the $(B+iJ)_1$-plane.}
  \label{fig:BJquin}
\end{figure}
\fi

The first striking feature of figure \ref{fig:BJquin} is that not all
values of $J_1$ are allowed. The sphere is mapped to the striped region
and so $J_1$ acquires a minimum value of $\ff12\cot\ff\pi5$. This fits
in with the idea of a minimum allowed size as we saw for the circle.
Any conformal field theory in this class may be associated with a \CY\
manifold of a size scale no less than the order of the Planck length.
The shaded area in figure \ref{fig:BJquin} represents the area in
which the instanton sum for the \CY\ nonlinear \sm\ converges.
Depending on one's taste, one might wish to declare this to set the
minimum length and then say that the bottom region must be describe in
terms of Landau-Ginzburg orbifolds.

A very important point to note is the following. The striped region in
figure \ref{fig:BJquin} may at first sight look similar to a
fundamental $Sl(2,\Z)$ region in the upper half plane as we had for
the torus example in section \ref{s:torus}. Sadly the region in figure
\ref{fig:BJquin} {\em cannot\/} be obtained this way. One may map this
region by a transcendental function to another region which may be
written in the form of a fundamental region \cite{CDGP:} but as this
map is not one-to-one it is not clear what purpose this can serve for
a model of the moduli space. The moduli spaces for the torus and for
K3 surfaces appeared naturally in the form of a Teichm\"uller space
divided by some modular group. It appears that for more generic $N$=2
theories such a description is no longer valid. This should not
dishearten the reader however --- we were still able to find
$\cM_A(X)$ without such a description.


\section{Phases}	\label{s:phase}

In the last section we looked at the simplest case of a moduli space
for a nontrivial case $d=3$. It was found that the space naturally
divided into two phases. In this section we will discuss the more
general picture.

\subsection{Another Example}  \label{ss:eg2}

To facilitate the discussion it will prove useful to run through
another example. This example was first analyzed in \cite{CDFKM:I}
(see also \cite{HKTY:}).
The idea is very similar to that of the quintic hypersurface except
this time we begin with a {\em weighted\/} projective space
$\P^4_{\{2,2,2,1,1\}}$. This is the space with homogeneous coordinates
$[x_0,\ldots,x_4]$, omitting $[0,0,0,0,0]$, where we identify
\begin{equation}
  [x_0,x_1,x_2,x_3,x_4]\cong[\lambda^2x_0,\lambda^2x_1,
	\lambda^2x_2,\lambda x_3,\lambda x_4],
\end{equation}
where $\lambda\in\C^*$. The desired \CY\ hypersurface is given by
\begin{equation}
f=x_0^4+x_1^4+x_2^4+x_3^8+x_4^8=0.
\end{equation}
This hypersurface is not smooth. There is a $\Z_2$ singularity along
the surface $[x_0,x_1,x_2,0,0]$ in $\P^4_{\{2,2,2,1,1\}}$ which
intersects our hypersurface along a curve. At each point on the curve,
this singularity is
essentially the same as the the one studied in section \ref{s:K3t}.
Thus we may blow each point of the curve up to get $\P^1$ to smooth
the space. The resulting exceptional divisor, $E$, will be a surface
(basically the old curve$\times\P^1$).
This smooth \CY\ manifold, containing the surface $E$, will be our
smooth model for $X$. The hyperplane in $\P^4_{\{2,2,2,1,1\}}$
produces another divisor in $X$ which we call $F$. The divisors $E$
and $F$ are associated to two linearly independent elements of
$H^2(X)$. In fact $h^{1,1}(X)=2$ and these elements form a basis.
Therefore $\cM_A(X)$ is now two-dimensional.

For Witten's linear \sm\ we now want $s=2$ and so we need two $U(1)$
charges and on complexification have two $\C^*$ actions to consider on
the target space. We assign the following charges:
\begin{equation}
\begin{tabular}{|c|c|c|}
  \hline
  $\phi_i$&$Q^{(1)}_i$&$Q^{(2)}_i$\\
  \hline
  $x_0$ & 0 & 1 \\
  $x_1$ & 0 & 1 \\
  $x_2$ & 0 & 1 \\
  $x_3$ & 1 & 0 \\
  $x_4$ & 1 & 0 \\
  $s$ & $-2$ & 1 \\
  $p$ & 0 & $-4$ \\
  \hline
\end{tabular}
\end{equation}
and have the following superpotential
\begin{equation}
  W = p(x_0^4+x_1^4+x_2^4+s^4x_3^8+s^4x_4^8+\ldots),
\end{equation}
where $\ldots$ represents other term that may be added (with arbitrary
coefficients) respecting the quasi-homogeneity of the equation.

The $D$-terms are now fixed by the equations of motion as
\begin{equation}
\eqalign{D_1&=|x_3|^2+|x_4|^2-2|s|^2-r_1,\cr
  D_2&=|x_0|^2+|x_1|^2+|x_2|^2+|s|^2-4|p|^2-r_2.\cr}
\end{equation}
Recall that for the example in section \ref{ss:eg1} we found that
fields which were sections of $\O(n)$ vanished when $n\leq0$ even
though one might initially expect this only to be the case when
$n<0$. The $n=0$ case should be checked by the transversality
condition. We assume this to be the case below except where noted.

We obtain four phases:
\begin{enumerate}
\item $r_1<0$ and $r_1+2r_2<0$: \hfil\break
  In this case $n_1\leq\ff12n_2$ and $n_2\leq0$ by the condition $\sum_l
n_lr_l\geq0$. The fields $x_0,x_1,x_2$ are sections of $\O(n_2)$ and the
fields $x_3,x_4$ are sections of $\O(n_1)$. Therefore they all vanish.
Taking the $D$-terms to vanish to solve $U=0$ forces a non-zero value
for $s$ and $p$ which may be fixed by the $(\C^*)^2$ gauge group. Thus
the target space is a point. There is a residual $\Z_8$ symmetry left
over so that our theory is actually a Landau-Ginzburg theory with
target space $\C^5/\Z_8$.
\item $r_1<0$ and $r_1+2r_2>0$: \hfil\break
  In this case $n_2\geq0$ which forces $p=0$. Solving for $U=0$, the
$D$-terms force $s$ to be non-zero and that $x_0,\ldots,x_4$ cannot
simultaneously vanish. We may use one $\C^*$ symmetry to fix $s$. The
other one may be used to turn $x_0,\ldots,x_4$ into the homogeneous
coordinates of $\P^4_{\{2,2,2,1,1\}}$. The target space is thus the
hypersurface $f=0$ in $\P^4_{\{2,2,2,1,1\}}$. We call this the ``orbifold''
phase since this hypersurface has $\Z_2$ quotient singularities.
\item $r_1>0$ and $r_2<0$: \hfil\break
  Now $n_1\geq0$ forcing $s=0$ and $n_2\leq0$ forcing $x_0=x_1=x_2=0$.
Solving for $U=0$ forces $p$ to be nonzero and $x_3$ and $x_4$ to not
simultaneously vanish. One $\C^*$ may be used to fix $p$ and the other
to turn $x_3,x_4$ into homogeneous coordinates on $\P^1$. The target
space is therefore $\P^1$. This is not quite the full story however,
the fluctuations of the fields $x_0,x_1,x_2,s$ are massless and governed by a
quartic superpotential. Fixing $p$ leaves a residual $\Z_4$ symmetry
meaning that that fields $x_0,x_1,x_2,s$ live in the space
$\C^4/\Z_4$. This phase is a hybrid-like phase consisting of a
Landau-Ginzburg (orbifold) bundle over $\P^1$. We call this the ``$\P^1$
phase''.
\item $r_1>0$ and $r_2>0$: \hfil\break
  This case is a little more complicated and we have to be more
careful concerning our assertion that sections
of $\O(m)$ are forced to vanish when $m\leq 0$. We have $n_2\geq0$
which will force $p=0$ when $n_2>0$ and we will find that $p=0$ by
transversality when $n_2=0$.
When $n_1>0$ we will have $s=0$ but we
need to exercise more care when $n_1=0$. Solving for $U=0$ we find
that $x_3,x_4$ cannot simultaneously vanish and that
$x_0,x_1,x_2,s$ cannot simultaneously vanish. First assume that
$s\neq0$. We may then use one of the $\C^*$-actions to fix $s$.
Let the other $\C^*$ make $x_0,\ldots,x_4$ into homogeneous coordinates.
Our target
space is now very similar to that of the hypersurface in
$\P^4_{\{2,2,2,1,1\}}$ except that we are missing the points where
$x_3=x_4=0$ and $x_0^4+x_1^4+x_2^4=0$. Now let $s=0$. Now we have
$x_0^4+x_1^4+x_2^4=0$ and $x_3$ and $x_4$ may be any value except both
zero. Use one of the $\C^*$-actions to turn $x_3,x_4$ into the
homogeneous coordinates of $\P^1$ and the other $\C^*$-action turn
$x_0,x_1,x_2$ into homogeneous coordinates. What we have is a
description of the smooth, blown-up \CY\ manifold described above ---
the singular point set has been replaced by an exceptional divisor
consisting of a curve times $\P^1$. This is the ``\CY'' phase.
\end{enumerate}

This produces a ``phase diagram'' which we show in figure
\ref{fig:ph}.

\iffigs
\begin{figure}[t]
  \centerline{\epsfxsize=10cm\epsfbox{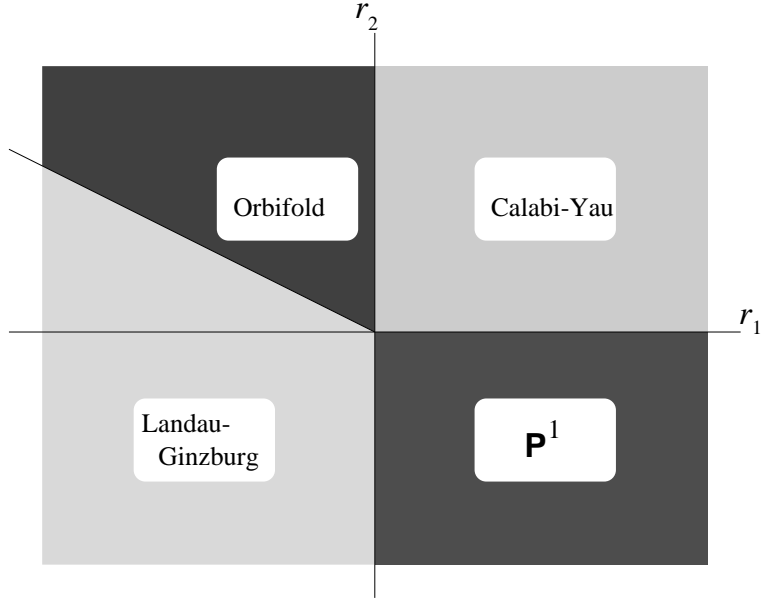}}
  \caption{Phase diagram for $h^{1,1}=2$ example.}
  \label{fig:ph}
\end{figure}
\fi

The instantons which appear in each sector can now appear in various
forms. The cases are given as follows:
\begin{enumerate}
\item For the Landau-Ginzburg orbifold we have only twist-fields.
\item For the orbifold phase we have two kinds. Firstly we have
rational-curve type instantons (together with their small-scale
supplements). Secondly we have twist-field instantons confined to the
region around the $\Z_2$-quotient singularity.
\item For the $\P^1$ phase we have rational curve type instantons ---
the rational curve being $\P^1$ itself! We also have twist-field
instantons in the Landau-Ginzburg fibre since this has a $\Z_4$
quotient singularity.
\item For the \CY\ phase we have rational curve instantons.
\end{enumerate}

Now let us consider the mirror, $Y$,  of this example. This is
obtained by dividing $X$ by the group $(\Z_4)^3$ consisting of elements
\begin{equation}
  g\colon(x_0,\ldots,x_4)\mapsto
	(e^{2\pi is_0}x_0,\ldots,e^{2\pi is_4}x_4),
\end{equation}
where $4s_0,4s_1,4s_2,8s_3,8s_4$ and $s_0+\ldots+s_4$ are all integers.
The general form of the defining equation for $Y$ is then
\begin{equation}
  f=x_0^4+x_1^4+x_2^4+x_3^8+x_4^8-8\psi x_0x_0x_0x_0x_0-
	2\phi x_3^4x_4^4,	\label{eq:fY2}
\end{equation}
where $\phi$ and $\psi$ are complex parameters that vary the complex
structure of $Y$. This is in agreement with our expectation that
$h^{2,1}(Y)=2$. To map the space spanned by $\phi$ and $\psi$ to the
moduli space of linear \sm s for $X$ we need to find special points in
these spaces. We first consider the values of $\psi$ and $\phi$ for
which the hypersurface (\ref{eq:fY2}) becomes singular. A little
algebra shows that $\partial f/\partial x_i=0$ for all $x_i$ admits
nontrivial solution when $\phi^2=1$ or $(\phi+8\psi^4)^2=1$. With some
foresight let us introduce the variables
\begin{equation}
  \eqalign{\rho_1 &= \ff1{2\pi}\log|4\phi^2|,\cr
  \rho_2 &= \ff1{2\pi}\log\left|\frac{2^{11}\psi^4}{\phi}\right|.\cr}
	\label{eq:rhos}
\end{equation}
We may now map the singular points in our moduli space into the
$\rho_1,\rho_2$ plane. That is we find the values for $\rho_1$ and
$\rho_2$ for which there can be a $\phi$ and $\psi$ which give a
singular $Y$. The result is shown in figure \ref{fig:ph2}

\iffigs
\begin{figure}[t]
  \centerline{\epsfxsize=10cm\epsfbox{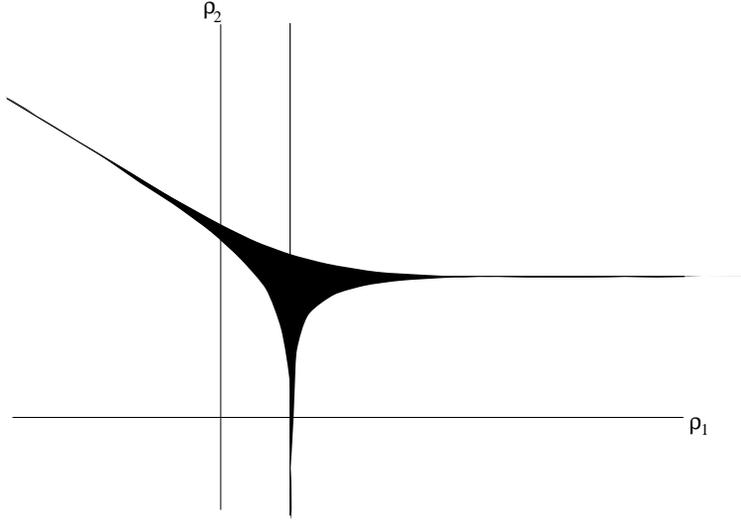}}
  \caption{Singular locus for $h^{1,1}=2$ example.}
  \label{fig:ph2}
\end{figure}
\fi

Now imagine that we ``zoom out'' infinitely far from figure
\ref{fig:ph2}. It should be easy to see that the shaded region will
become precisely the phase boundaries in figure \ref{fig:ph}.
In fact, the variables in (\ref{eq:rhos}) were chosen precisely so
that we may make the identification $r_1=\rho_1$ and $r_2=\rho_2$.
The fact
that the asymptotes of the region in figure \ref{fig:ph2} are parallel
to, rather than along, lines passing through the origin comes from the
one-loop renormalization effects of $r_l$ --- just as we had a $5\log5$
shift in section \ref{ss:eg1}.
Thus assuming a simple map between the B-model moduli space
coordinates and the linear \sm\ coordinates as we had in
(\ref{eq:zpsi}), we obtain
\begin{equation}
  \eqalign{z_1 &= \frac1{4\phi^2},\cr
  z_2 &= \frac{\phi}{2^{11}\psi^4}.\cr}
	\label{eq:zpsi2}
\end{equation}
The reader is referred to \cite{MP:inst} for the complete proof that
this is indeed the correct map.

Again we may calculate the Picard-Fuch's equation for this
two-parameter example \cite{CDFKM:I}. Again (\ref{eq:q=z}) uniquely
determines the relationships
\begin{equation}
\eqalign{q_1&=z_1+2z_1^2+48z_1z_2+5z_1^3+7560z_1z_2^2+\ldots\cr
q_2&=z_2-z_1z_2+104z_2^2-z_1^2z_2-56z_1z_2^2+15188z_2^3+\ldots\cr}
\end{equation}
This is a convergent power series in the Calabi-Yau phase.

As before each phase leads to some notion of a series for correlation
functions which will be convergent in some region. In the example
studied in section \ref{ss:eg1} we found that any generic point in the
moduli space lay in the region of convergence of one of the power
series. This is no longer the case in this two parameter example. The
shaded region in figure \ref{fig:ph2} marks where none of the power
series converge. In this sense, the phase picture does not
cover the entire moduli space although it will cover most of a circle
with center $r_1=r_2=0$ in the limit of infinite radius.

As we did in section \ref{ss:eg1} we may analytically continue the
$q_l$'s into the other three phases. This calculation was done in
\cite{me:min-d}. Doing so one obtains the following bound for the
entire moduli space
\begin{equation}
\fbox{$J_1\geq0.$}
\end{equation}
This is a very important result. It means that the algebraic curves in
$X$ whose areas are measured by $J_1$ (i.e., the rational curves in
the exceptional divisor) may shrink down to zero size. {\bf We have
broken the constraint of a minimum length.}

The reader may argue that we analytically continued $q_l$ beyond the
\CY\ phase and so we may artificially have been able to reach zero
size. Actually in this case, $J_1$ can take on arbitrarily small
values while one remains in the \CY\ phase \cite{me:min-d}. Thus one
really is forced to reject a notion of minimum length in this example.

It is worth noting that general arguments have been made which appear
to prove a universal notion of minimum length (such as
\cite{KPP:min}). These arguments appear to rely on the assumption that
space-time is locally flat, i.e., one believes the
differential geometry view. Unless we can concoct some alternative
view of small distances this will appear to show that differential
geometry is seriously misleading at small scales!

\subsection{More general cases}

The examples studied so far show many of the aspects of the global
structure of the moduli space --- $\cM_B(X)$ is rather dull and can be
explained classically while $\cM_A(X)$ forces a ``phase'' description
upon us and forces us to rethink our notions of geometry at small
scales. Let try to understand the more general picture of $\cM_A(X)$.

One of the phase transitions for the example studied in section
\ref{ss:eg2} was between an orbifold and a \CY\ manifold. It was
precisely the blow-up of section \ref{s:K3t}. All of the phase
transitions mentioned thus far are actually of this type.

Let us return to the example of the quintic threefold of section
\ref{ss:eg1}. The two phases here were a smooth \CY\ manifold and an
Landau-Ginzburg theory in $\C^5/\Z_5$. The natural question to ask is
if we can blow-up the quotient singularity $\C^5/\Z_5$. The answer is
yes, and the procedure is very similar to that of blowing up the
$\C^2/\Z_2$ singularity. In the latter case we used the line bundle
over $\P^1$ with first Chern class $-2$. For $\C^5/\Z_5$ we use the line
bundle over $\P^4$ with first Chern class $-5$. That is, the
exceptional divisor is $\P^4$ --- but $\P^4$ is precisely the ambient
space for the \CY\ phase. This leads us to the following general
picture. Define $X$ as the critical point set of some function $W$ on
some (non-compact) space, $V$. By blowing up and down quotient
singularities in $V$ we induce the phase transitions in $X$.

The singularity $\C^5/\Z_8$ is blown up using an exceptional divisor
with two irreducible components. Having each of these components blown
up or down leads to the 4 phases. For example, the first component is
$\P^1\times\C^3$. Blowing up the Landau-Ginzburg phase using this
component gives the $\P^1$ phase. This still has $\Z_4$ singularities
all along this $\P^1$. The second component is $\P^4$. This component
resolves the $\Z_4$ singularities in each fibre. The restriction to
the critical point set of $W$ in each fibre forms a quartic constraint. A
quartic in constraint in $\P^4$ is a K3 surface. Thus this latter phase
is a fibre bundle over $\P^1$ with generically a K3 fibre. This is one
description of the smooth \CY\ manifold \cite{CDFKM:I}. We may also
reach the smooth phase by blowing up along $\P^4$ first to obtain the
orbifold phase and then along $\P_1\times\C^3$ to resolve the
orbifold.

Have we constructed the general picture by considering blow-ups of
quotient singularities? The
answer to this question is no. We should then ask the question as to
whether there is any transformation that might give the general
picture. The answer that appears to be the case
is that we consider ``birational'' transformations. Birational
transformations occur very naturally in algebraic geometry and thus we
shouldn't be too surprised that they will be natural objects in $N$=2
theories. Two algebraic varieties $X_1$ and $X_2$ are birationally
equivalent if one can find open subsets $U_1\subset X_1$ and
$U_2\subset X_2$ such that the set of functions in $U_1$ is isomorphic
to the set of functions in $U_2$ (see \cite{Hartshorne:} for a more
careful definition). An example of birationally equivalent pairs are
given by $X_1$ being a blow-up of $X_2$.

Another example of birational equivalence can be provided still from
quotient singularities. This stems from the fact that the process of
blowing-up a quotient singularity need not be a unique process.
Suppose we have a singular space $X_0$ which may be smoothed by
blowing-up into two
topologically distinct smooth spaces $X_1$ and $X_2$. These two smooth
spaces are then birationally equivalent.

In \cite{AGM:I,AGM:II} an example with $h^{1,1}(X)=5$ was studied.
This requires toric geometry which takes us beyond the scope of these
lectures so we will not provide details here. There are 100 phases in
all for this model of which 5 consist of smooth \CY\ manifolds. These
5 manifolds are all the possible resolutions of an orbifold which
provides one of the other phases.

One may also picture a direct transition between these manifolds in
the form of a ``flop'' as we now explain. We can raise our example of
a $\C^2/\Z_2$ quotient singularity of section \ref{s:K3t} to one
higher dimension by considering the space $xy-wz=0$ in $\C^4$ with
coordinates $(x,y,z,w)$. Transversality tells us that this
hypersurface has an isolated singularity at $(0,0,0,0)$. It cannot be
written as a quotient singularity. Now consider the space
\begin{equation}
  \O(-1,-1) = \{[a,b],(x,y,z,w)\in\P^1\times\C^4;\, az=bx, ay=bw\}.
	\label{eq:flop1}
\end{equation}
This smooths the singularity, in a manner similar to blow-ups
discussed earlier, replacing the singular point by $\P^1$. The space
\begin{equation}
  \O(-1,-1) = \{[a,b],(x,y,z,w)\in\P^1\times\C^4;\, aw=bx, ay=bz\}
	\label{eq:flop2}
\end{equation}
does pretty much the same thing. The only difference between
(\ref{eq:flop1}) and (\ref{eq:flop2}) is the way in which the $\P^1$
is inserted. The two smooth spaces produced are said to differ by a
flop.
This local picture may be fitted into more complicated global
geometries.

These five smooth models for $X$ occupy adjacent phases (since they
are related by flops) in the phase
picture and so we may consider a path of conformal field theories
passing from one to the other. So long as this path is generic, it
will not hit the singularity lying in the complex codimension one
``phase boundary'' and so this transition is perfectly smooth from the
conformal field theory point of view. This shows how string theory can
give rise to a ``smooth topology change'' in the target space so long
as the two topological spaces are birationally equivalent as algebraic
varieties.

If the reader was not convinced thus far that algebraic geometry was
superior to differential geometry for our purposes then this last
point must surely convince them. The natural equivalence class in
differential geometry is that of diffeomorphic equivalence which is a
class stronger than topological equivalence. String theory happily
combines different classes smoothly into the same moduli space ---
string theory is oblivious to such distinctions! However, every phase
in a given moduli space belongs to single birational
equivalence class.

There are many more possibilities of birational transformations than
those provided by resolving quotient singularities and flops. One of
the more exotic ones is the ``exoflop'' of \cite{AGM:II}. In the case
of the flop, one $\P^1$ is transformed into another one within a \CY\
manifold. In the the case of an exoflop, a rational curve is taken
from within the \CY\ manifold to one glued onto the outside.

The complete picture of possible phases is far from complete. The most
general construction for models for which one understands the linear
\sm\ is that provided by \cite{Boris:m}. These are explored to some
extent \cite{AG:gmi} although the full range of possibilities have yet
to be classified. One interesting point discussed in \cite{AG:gmi} is
that the phase diagrams need not contain any smooth \CY\ phase. This
really is quite natural given that there is nothing particularly
special about the smooth \CY\ phase --- it has no more reason to exist
than any of the other phases. The reason we tend to like \CY\
manifolds rather than the other phases is because we tend to be happier with
differential geometry which only works properly in this phase.

Having said that we should rid our minds of any bias towards the \CY\
manifold phase we should note some rather interesting curiosities which
occur when we try to analyze the whole moduli space in terms of these smooth
phases. Let us consider the five parameter example of \cite{AGM:II}
which has 100 phases of which 5 are \CY\ manifolds. When we do the
analytic continuation analysis to map out the entire moduli space in
terms of one phase, which \CY\ phase should we begin with?
The surprising answer is that is doesn't actually matter as we now explain.

Suppose we begin with one of the topologies for $X$ which we denote
$X_1$. Thus we have some phase in the moduli space containing the
large radius limit of $X_1$ where we put
$z_1=\ldots=z_5=q_1=\ldots=q_5=0$. Now let
$X_2$ be another topology which is obtained by flopping $X_1$. This
flop will correspond to a rational curve $C_1\subset X_1$ and
another rational curve $C_2\subset X_2$ which arise in the form of
equations (\ref{eq:flop1}) and (\ref{eq:flop2}). We may chose $q_5$ such
that
\begin{equation}
  \hbox{Area}(C_1)=J_5.
\end{equation}
Now we consider the flop process. To make things simple we should make
every rational curve in $X_1$ have infinite area except for
$C_1$. This may be achieved by setting $q_1=\ldots=q_4=0$. Now we
analyze the Picard-Fuch's equation to obtain $q_5$ in terms of
$z_5$. The result is \cite{AGM:sd}
\begin{equation}
  q_5=z_5.
\end{equation}
Analysis of the phase picture shows that if we introduce coordinates
$z^\prime_1,\ldots,z^\prime_5$ for a patch of coordinates with origin
at the large radius limit of $X_2$ then $z^\prime_5=(z_5)^{-1}$. We
also have $q^\prime_5=z^\prime_5$. This means that the analytic
continuation between these two phases simply asserts that the area of
$C_1$ is minus the area of $C_2$. Thus we naturally identify a topology
with a certain conformal field theory by demanding that all areas be
positive.

It appears to be the case \cite{AGM:sd} that the following now
applies in the case that there is at least one \CY\ phase. Given any
conformal field theory we may associate some \CY\ target
space topology (possibly by analytic continuation) in which all areas
are non-negative. That is to say when we go from a phase picture in
terms of $r_l$'s which covers the whole space $\R^s$ and then remap this
space into the space of $J_l$'s then only points within (or on the
boundary of) cones corresponding to smooth \CY\ phases are
covered. This is shown schematically in figure \ref{fig:Ka} for a
hypothetical example with $s=2$ and five phases $X_1,\ldots,X_2$ for
which only $X_1$ and $X_2$ are smooth \CY\ manifolds.

\iffigs
\begin{figure}[t]
  \centerline{\epsfxsize=12cm\epsfbox{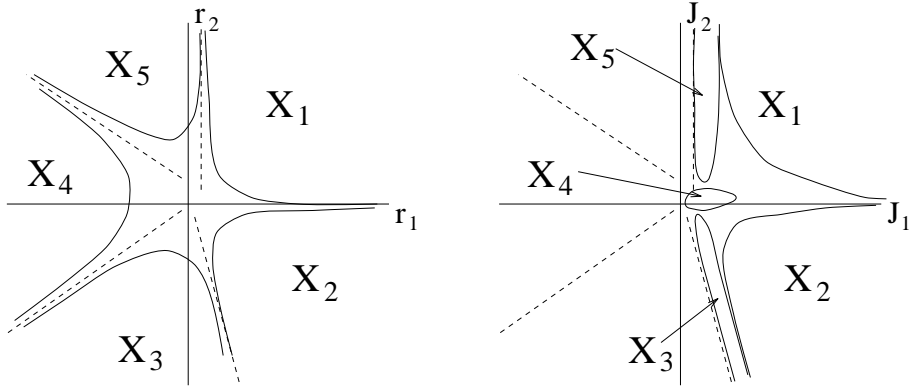}}
  \caption{Moving between $r_l$ and $J_l$.}
  \label{fig:Ka}
\end{figure}
\fi


\section{Conclusions}	\label{s:conc}

We hope the reader is convinced that algebraic geometry together with
mirror symmetry is a very
useful tool for analyzing the moduli space of $N$=2 theories. Although
sticking to metric-based ideas seems to work without any problems for
some models, it
appears to face severe short-comings in the generic case. In
particular, when there are instantons one tends to have phases since
one will have instanton sums which are only convergent in a certain
region of the moduli space. When this happens the other phases appear
to have an equal right to be taken as a model for the target space and
then one necessarily needs to discuss singular spaces.

Where metric-based ideas may work, in the case $d=3$, appears to be
restricted to the few cases where $h^{2,0}>0$ (which implies that
$h^{1,0}>0$ which in turn implies that the target space has a
continuous isometry). Although we didn't discuss these examples in
detail, the interested reader may try to apply the methods of section
\ref{s:torus} and \ref{s:K3t} for these cases.

The methods used in section \ref{s:d=3} have certain shortcomings
which may not have been completely apparent.
Firstly there may be some subtleties introduced into the phase picture
if there are some symmetries of the defining equation that are not of
the most na{\"\i}ve kind. We refer the reader to \cite{CFKM:II} for
such an example.
Another, more severe, problem is that not
all of the dimensions of the moduli space may be parametrized by the
linear \sm. That is, we may only be able to write down models based
on a gauge group $U(1)^s$ where $s<h^{1,1}(X)$. The mirror to this
statement is the fact that not all deformations of a complete intersection
can, in general, be written as deformations of the defining equations
\cite{GH:poly}. We do not have any techniques at our disposal to
address the complete moduli spaces of such objects at this point in
time.

We should also repeat that some $N$=2 conformal field theories cannot
be put in the form of a gauged $U(1)$ linear \sm\ at all. This is
equivalent to the statement in geometry that not all \CY\ manifolds
can be written as complete intersections in toric varieties. One may
try to extend methods to nonabelian gauge groups. Although such models
were discussed in \cite{W:phase}, since the mirror map construction
for such models is not yet understood, we may not apply most of the
methods discussed in this paper. Also there is no reason to believe
that these nonabelian groups will exhaust all the $N$=2 theories.

Although the moduli space of $N$=2 theories is far better understood
now than people imagined a few years ago it appears that there is
still much left to discover.


\section*{Acknowledgements}

It is a pleasure to thank B. Greene, D. Morrison and R. Plesser
for many useful conversations.
The work of the author is supported by a grant from the National
Science Foundation.


\end{document}